\documentstyle[psfig,emulateapj]{article}
\def\etal   {{et~al.}\ }

\def\kms{{\rm\,km/s}}
\def\msun{{\rm\,M_\odot}}

\def\vol#1  {{{#1}{\rm,}\ }}

\def\etal{et al.\ }

\def\clock{\count0=\time \divide\count0 by 60
     \count1=\count0 \multiply\count1 by -60 \advance\count1 by \time
     \number\count0:\ifnum\count1<10{0\number\count1}\else\number\count1\fi}

\font\eightrm=cmr8 scaled \magstep0
\def\fig #1, #2, #3, #4, #5, #6 {
\topinsert
\smallskip
\centerline{\psfig{figure=#1,height=#2 in,width=#3 in,angle=#4}}
\medskip
{\vskip #5 cm\leftskip2.5em \parindent=0pt {\eightrm #6 }}
\endinsert}

\begin{document}
\title{Gaussian Peaks and Clusters of Galaxies}
\author{Renyue Cen\altaffilmark{1}}
\altaffiltext{1} {Princeton University Observatory, Princeton University, Princeton, NJ 08544; cen@astro.princeton.edu}

\begin{abstract}

We develop and test a method to compute 
mass and auto-correlation functions of rich clusters 
of galaxies from linear density fluctuations,
based on the formalism of Gaussian peaks (Bardeen \etal 1986).
The essential, new ingredient in the current approach
is a simultaneous and unique fixture of 
the {\it size} of the smoothing window for the density field, $r_f$,
and the {\it critical height}
for collapse of a density peak, $\delta_c$,
for a given cluster mass
(enclosed within the sphere of a given radius rather than the virial
radius, which is hard to measure observationally).
Of these two parameters, $r_f$
depends on both the mass of the cluster
in question and $\Omega$,
whereas $\delta_c$
is a function of only $\Omega$ and $\Lambda$.
These two parameters have formerly 
been treated as adjustable and approximate parameters.
Thus, for the first time,
the Gaussian Peak Method (GPM) becomes unambiguous,
and more importantly, accurate, as is shown here.

We apply this method to constrain all variants 
of the Gaussian cold dark matter (CDM) cosmological model
using the observed abundance of local rich clusters of galaxies
and the microwave background temperature fluctuations
observed by COBE.
The combined constraint 
fixes the power spectrum of any model to $\sim 10\%$ accuracy
in both the shape and overall amplitude.
To set the context for analyzing CDM models,
we choose six representative models of current interest,
including
an $\Omega_0=1$ tilted cold dark matter model,
a mixed hot and cold dark matter model 
with $20\%$ mass in neutrinos,
two lower density open models with $\Omega_0=0.25$ and $\Omega_0=0.40$,
and 
two lower density flat models with $(\Omega_0=0.25,\Lambda_0=0.75)$ 
and $(\Omega_0=0.40,\Lambda_0=0.60)$.
This suite of CDM models should bracket
any CDM model that is currently viable.
The parameters of 
all these models are also consistent with a set of other constraints,
including the Hubble constant, the age of the universe
and the light-element nucleosynthesis with
$\Omega_b$ chosen to maximize the viability of each model
with respect to the observed gas fraction in X-ray clusters.

\end{abstract}



\section{Introduction}

Clusters of galaxies are cosmologically important because
they contain vitally important information on 
scales from a few megaparsecs to several hundred megaparsecs,
and provide fossil evidence
for some of the basic cosmological parameters
(Richstone, Loeb, \& Turner 1992;
Bahcall, Fan, \& Cen 1997).
The fact that they are among the most luminous objects
in the universe renders them an effective
and economical tracer of the large-scale
structure, not only of the local universe (Bahcall 1988) 
but also of the universe at moderate-to-high redshift.

Perhaps more interesting is the fact that 
clusters of galaxies are intrinsically rare 
with typical separations of $\sim 50h^{-1}\;$Mpc at $z\sim 0$
and seemingly rarer at high redshift
(Luppino \& Gioia 1995;
Carlberg \etal 1996;
Postman \etal 1996).
Their rarity is traceable to the fact that
they only form at the rare, high peaks in the initial density field.
Since the mass in a sphere of radius $10h^{-1}$Mpc
roughly corresponds to the mass of a rich cluster like the Coma cluster,
the abundance of clusters of galaxies 
(i.e., the cluster mass function, Bahcall \& Cen 1992)
provides a sensitive test of the amplitude of the density 
fluctuations on that scale and
places one of the most stringent constraints on cosmological 
models to date
(\cite{pdj89};
\cite{ha91}; 
\cite{bc92}; 
\cite{ob92}; 
\cite{bs93};
\cite{wef93}, WEF henceforth;
Viana \& Liddle 1996;
Eke, Cole \& Frenk 1996, ECF henceforth; 
Pen 1996)

The spatial distribution of clusters of galaxies 
provides complementary information for cosmological models.
A widely used statistic for clusters of galaxies
is the two-point auto-correlation function. 
Earlier pioneering work (Bahcall \& Soneira 1983; Klypin \& Kopylov 1983)
has met with dramatic improvements in recent years
thanks to larger and/or new cluster samples that
have become available
(Postman, Huchra, \& Geller 1992;
Nichol \etal 1992; 
Dalton \etal 1992,1994; 
Romer \etal 1994;  
Croft \etal 1997   
).
Comparing with cosmological models clearly show
that the two-point correlation function of clusters of galaxies
provides a strong test on cosmological models
on scales from several tens to several hundred megaparsecs
(\cite{bbe87}; Bahcall \& Cen 1992;
Mann, Heavens, \&  Peacock 1993;
Holtzman \& Primack 1993;
Croft \& Efstathiou 1994;
Borgani \etal 1995).
In addition, the recent studies 
of superclusters and supervoids by Einasto \etal (1997a,b,c)
show a very intriguing property that 
the correlation function 
of rich clusters appears to be oscillatory on large scales.
If confirmed, this would challenge most models.

Hence, the combination
of cluster mass and correlation functions
provides a critical constraint on cosmological 
models on scales $\ge 10~h^{-1}$Mpc.
While uncertainties remain in the 
current clustering analyses as well as the abundance
of observed clusters
due chiefly to still limited cluster sample sizes typically
of order of a few hundred clusters, large 
surveys underway such as the Sloan Digital Sky Survey (SDSS; 
Knapp, Lupton, \& Strauss 1996) and
2dF galaxy redshift survey (Colless 1998)
should provide much more accurate determinations of both.

The groundwork for the gravitational instability picture
of cluster formation was laid down 
more than two decades ago 
(Gunn \& Gott 1972).
In the context of Gaussian cosmological models,
Kaiser (1984), in a classic paper, 
put forth the ``biased" structure formation mechanism,
where clusters of galaxies were predicted (correctly) to form
at high peaks of the density field
to explain the enhanced correlation of Abell clusters
over that of galaxies.
This idea was subsequently extended to objects on other scales
including galaxies and 
the properties of linear Gaussian density fields
were worked out in exquisite detail (Peacock \& Heavens 1985; 
Bardeen \etal 1986, BBKS hereafter).

While alternatives exist (e.g.,
Zel'dovich 1980; Vilenkin 1981,1985;
Turok 1989;
Barriola \& Vilenkin 1989; Bennett \& Rhie 1990),
a Gaussian model is simple and attractive (largely because of it)
in that all its properties can be fully specified by one single
function, the power spectrum of its density fluctuations.
Moreover, it is predicted that
random quantum fluctuations generated in the early universe 
naturally produced Gaussian density fluctuations,
whose scales were then stretched to the scales of cosmological interest 
by inflation (Guth \& Pi 1982; 
Albrecht \& Steinhardt 1982;
Linde 1982;
Bardeen, Steinhardt \& Turner 1983).
Furthermore, observations of large-scale structure and microwave background
fluctuations appear to favor a Gaussian picture 
(Vogeley \etal 1994;
Baugh, Gaztanaga, \& Efstathiou 1995;
Kogut \etal 1996;
Colley, Gott, \& Park 1996;
Protogeros \& Weinberg 1997;
Colley 1997).
So motivated, 
the present study will focus on the family of Gaussian CDM models.
The reader is referred to Cen (1997c) for a discussion of the cluster 
correlation function in non-Gaussian models.
We will employ 
the formalism of BBKS of Gaussian density field 
to devise an analytic method
that can be used to directly compute the
mass and correlation functions of clusters of galaxies.
The needed input are: $P_k$ (the power spectrum),
$\Omega_0$ and $\Lambda_0$.
The method developed is calibrated and
its accuracy checked by a large set of N-body simulations.

The motivation for having such an analytic 
method is not only of an economical consideration (fast speed
and much larger parameter space coverage possible)
but also a necessity,  especially for studying very rich clusters.
For example, for clusters of mean separation of $200h^{-1}$Mpc
(about richness 3 and above; Bahcall \& Cen 1993, BC henceforth),
a simulation box of size $1170h^{-1}$Mpc on a side
would contain 200 such clusters, a number which 
may be required for reasonably sound statistical calculations.
Assuming that the mass of such a cluster is $1.0\times 10^{15}h^{-1}\msun$
(approximately the mass of a richness 3 cluster; BC)
and one requires $500$ particles to claim an adequate 
resolution of the cluster, it demands a requisite particle mass
of $2.0\times 10^{12}h^{-1}\msun$.
This particle mass requirement dictates
that one discretize the whole simulation box
into $10^{8.3}\Omega_0$ particles 
($\Omega_0$ is the density parameter of a model).
Meantime, a minimum nominal spatial resolution of $0.5h^{-1}$Mpc
is needed to properly compute just the cluster masss within 
Abell radius of $1.5h^{-1}$Mpc, which translates to a
spatial dynamic range of 2340.
The combined spatial and mass resolution requirements
are formidable for either PM code, which requires a very large
mesh ($2340^3$) thus needs more than $57$GB of RAM to allow for such
a large simulation and hence is very expensive, if possible,
or an adaptive code such as P$^3$M (Efstathiou \etal 1985)
or TPM code (Xu 1995),
where CPU cost will be prohibitively large even if RAM permits.

The paper is organized as follows.
Descriptions of GPM for computing 
cluster mass function are presented in \S 2.1.
Descriptions of GPM for computing 
the cluster correlation function are presented in \S 2.2.
A calibration of Press-Schechter method 
using the fitted GPM parameters 
and some comparisons between GPM and Press-Schechter method
are presented in \S 2.3.
We discuss the various factors that affect the
cluster mass function in \S 3.
Detailed constraints by the 
local rich clusters and the COBE observations (Smoot \etal 1992)
on all CDM models are presented in \S 4.
A simple $\sigma_8-\Omega_0$ relation (with errorbars) for
CDM models is presented in \S 5,
derived from fitting to the observed local cluster abundance alone.
Conclusions are given in \S 6.

\section{Gaussian Peak Method for Clusters of Galaxies}

\subsection{Gaussian Peak Method for Cluster Mass Function}

It is convenient 
to define some  frequently used symbols first.
Hubble constant is $H=100h$km/s/Mpc.
$\Omega_0$ and $\Lambda_0$ are the density parameters due to
non-relativistic mater and cosmological constant, respectively,
at redshift $z=0$.
$\Omega_z$ and $\Lambda_z$ are the same parameters at redshift $z$.
$r_A$ is the comoving radius of a sphere in units of $h^{-1}$Mpc,
which in most times represents the Abell radius with value $1.5$.
$r_v$ is the virial radius in comoving $h^{-1}$Mpc.
$r_f$ is the radius of a smoothing window in comoving $h^{-1}$Mpc.
$M_A$ is the mass within a sphere of radius $r_A$ in units
of $h^{-1}\msun$.
$M_v$ is the mass within a sphere of radius $r_v$ (virial mass of a halo)
in units 
of $h^{-1}\msun$.
For formulae related to Gaussian density field
we will follow the notation of BBKS throughout this paper.

The cluster mass function
may be derived by relating the initial density peaks to the final
collapsed clusters,
provided that {\it peaks do not merge.}
Two pieces of observations suggest that
merger of initial density peaks of cluster size
be infrequent.
First, the typical separation between clusters of galaxies
is $\sim 100\;h^{-1}$Mpc, while the typical size of a cluster
is $\sim 1\;h^{-1}$Mpc.
Second, empirical evidence of matter fluctuations,
as indicated by observed galaxy number fluctuations 
(Davis \& Peebles 1983; Strauss \& Willick 1995),
suggests that the current nonlinear scale 
is $\sim 8\;h^{-1}$Mpc, which is just about the size of fluctuations
that collapse to form clusters of galaxies;
i.e., the majority of clusters of galaxies form at low redshift.
However, a more quantitative argument, that
merger rate should be small, can be made as follows.
Suppose that a cluster is moving at velocity $v_0$ at $z=0$,
then we can compute the total comoving distance that the cluster
has travelled in its entire lifetime as
\begin{equation}
d_{cm} = \int_0^{t_0} v_0 f(t) (1+z)dt,
\end{equation}
\noindent 
where $t_0$ is the current age of the universe, $f(t)$
is a function to describe the evolution of the (proper) peculiar
velocity of the cluster and the last term $(1+z)$ relates
comoving distance to proper distance.
To illustrate the point we will first use $\Omega_0=1$, which
gives the following simple relation:
\begin{equation}
d_{cm} = v_0 H_0.
\end{equation}
\noindent 
To arrive to the above relation we have used the following 
simple relations: $t=t_0 (1+z)^{-3/2}$, $t_0=2H_0^{-1}/3$,
$f(z)=(1+z)^{-1/2}$ (linear growth rate of proper peculiar
velocities; Peebles 1980).
For any reasonable model
clusters of galaxies
do not move at a speed (peculiar velocity) much higher than 
$\sim 1000$km/s at present 
(Cen, Bahcall \& Gramann 1994);
it can be obtained approximately in linear theory 
by integrating
a power spectrum, smoothed by an appropriate window,
to yield the total kinetic energy
(e.g., Suto, Cen, \& Ostriker 1992).
Note that some galaxy in a virialized cluster may move at a higher 
speed, but we are not considering such objects.
So, a cluster moving at $1000\kms$ today moves
a total comoving distance of $10h^{-1}$Mpc in an $\Omega_0=1$ universe.
The same cluster will move a longer distance in a 
low $\Omega_0$ universe, but not by a large factor.
An upper bound on $d_{cm}$ in such cases may be 
obtained by setting $\Omega_0=0$, in which case we have
$f(z)=(1+z)$, $t_0=H_0^{-1}$ and $t=t_0 (1+z)^{-1}$.
The upper bound is
\begin{equation}
d_{cm,ub} = v_0 H_0 (1+z_{max}),
\end{equation}
\noindent 
where $z_{max}$ is the maximum redshift to which $\Omega=0$ applies.
Let us make a simple, approximate 
estimate for a realistic lower bound by taking
$\Omega_0=0.2$, as follows.
For an $\Omega_0=0.2$ model, the redshift at which $\Omega=0.5$
is $3.0$, which we denote as $z_{max}$.
We treat the redshift range $z>z_{max}$ as an $\Omega=1$ model and
treat $z<z_{max}$ as an $\Omega=0$ model, which
gives a simple result:
\begin{equation}
d_{cm}=v_0 H_0 \left[(1+z_{max}) + 1\right].
\end{equation}
\noindent 
For $z_{max}=3.0$, $d_{cm}=5 v_0 H_0$.
Since velocity decayed from $z_{max}$ to $z=0$, 
a more reasonable upper bound on $v_0$ is 
$1000/(1+z_{max})$km/s.
This gives $12.5h^{-1}$Mpc for $v_0=1000$km/s and $z_{max}=3$.

Since one needs to collapse a sphere of $9.5\Omega_0^{-1/3}h^{-1}$Mpc in a
uniform density field to form a massive cluster
of mass $1\times 10^{15}h^{-1}\msun$,
i.e., cluster density peaks have to have a separation of at least
$\sim 20\Omega_0^{-1/3}h^{-1}$Mpc 
and more likely $\sim 50h^{-1}$Mpc (mean separation 
of rich clusters today),
it thus seems quite unlikely that a significant fraction
of any massive cluster peaks have merged by $z=0$.
This conclusion is, however, not in conflict with observations
that seem to show signs of recent and/or ongoing 
merger activity. 
In general, merger is an ever-going processes (at least in 
the past) in any plausible (i.e., a plausible
range in $\Omega_0$) hierarchical structure formation model.
But, these mergers or substructures seen in some clusters
are sub rich cluster scale mergers, i.e.,
sub-peaks within a large cluster scale peak
are in the process of merging, a result which is in fact
expected if clusters have been forming in the recent past 
in a hierarchical fashion.
To our knowledge, there is no major merger event 
of two massive clusters observed.
For example, in 55 Abell clusters catalogued by Dressler (1980),
there is no case of two massive clusters in the process of imminent
merging,
although there does seem
to have significant substructures in a significant
fraction of clusters in various optical studies (e.g.,
Geller \& Beers 1982; Dressler \& Shectman 1988;
West, Oemler, \& Dekel 1988).
X-ray observations also show a large fraction of clusters
with substructures 
(see Forman \& Jones 1994 for a review).
That being said, one needs to be extra cautious in interpreting
such sub cluster scale merger/substructure events, due to
unavoidable projection contaminations 
(see Cen 1997 for a thorough discussion of projection effects).

Having shown that merger should be infrequent,
the key link then is to relate a density peak of height 
\begin{equation}
\nu=F/\sigma_0
\end{equation}
\noindent 
to the final
mass of a cluster defined
within a fixed radius, say, the Abell radius $r_A$.
Here, $F$ is a density field smoothed by a window of size $r_f$
and $\sigma_0$ is the rms fluctuation of $F$.
Gaussian smoothing window (in Fourier space) 
\begin{equation}
W(k r_f) = \exp (-r_f^2 k^2/2),
\end{equation}
\noindent 
will be used
throughout this paper,
because it guarantees convergence of any spectral
moment integral with any plausible power spectrum.
Top-hat smoothing does not have this feature.
For the sake of definiteness and convenience in comparing with 
observations, we define a cluster mass, $M_A$,
as the mass in a sphere of {\it comoving}
Abell radius, $r_A=1.5\;h^{-1}$Mpc, in most cases.
Cluster mass defined otherwise will be noted in due course.
But the formalism developed here
should be applicable for any plausible radius.

For a spherical perturbation, 
the mean density within the virial radius (at redshift $z$)
in units of the global mean density (at redshift $z$)
can be parameterized by 
\begin{equation}
\bar\rho_v (\Omega_z,\Lambda_z) = 178 \Omega_z^{-0.57} C(\Omega_z,\Lambda_z).
\end{equation}
\noindent 
For $\Omega_z=1$ and $\Lambda_z=0$, 
it is well known that $\bar\rho_v = 178$, 
a result first derived by Gunn \& Gott (1972).
In equation (7)
$C(\Omega_z,\Lambda_z)$ (a function of both
$\Omega_z$ and $\Lambda_z$) has a value close to unity.
It has been shown that $C=1$ is a good approximation
for both $\Lambda_z=0$ model (e.g., Lacey \& Cole 1993)
and $\Lambda_z+\Omega_z=1$ model (e.g., ECF) 
for the range of $\Omega_z$ of interest ($0.1 < \Omega_z < 1.0$).
The mass within the comoving virial radius, $r_v$, at redshift $z$,
is therefore
\begin{eqnarray}
M_{v}(z)\hskip -0.3cm &=&\hskip -0.3cm{4\pi\over 3} r_v^3 \bar\rho_v(\Omega_z,\Lambda_z) \rho_{c} \Omega_0  \nonumber\\
\hskip -0.3cm&=&\hskip -0.3cm 2.058\times 10^{14} \Omega_z^{-0.57} \Omega_0 r_v^3 C(\Omega_z,\Lambda_z)
\end{eqnarray}
\noindent 
where $\rho_{c}$ is the critical density at $z=0$.
Since the spherical perturbation smoothed
by a Gaussian window of comoving radius $r_f$ is, by assumption,
just virialized at the redshift in question, we 
have another expression for $M_v$
as a function of $r_f$:
\begin{equation}
M_{v}(z) = (2\pi)^{3/2} r_f^3 \rho_{c} \Omega_0 
=4.347\times 10^{12}\Omega_0 r_f^3
\end{equation}

\noindent 
Next, we need to relate $M_v$ to $M_A$, 
the latter of which is observationally more obtainable.
We assume the following simple relationship:
\begin{equation}
M_{A} = M_v ({r_A\over r_v})^{3-\alpha}.
\end{equation}
\noindent 
This relation holds exactly, if the  density profile of the cluster
has the power-law form:
\begin{equation}
\rho(r) \propto r^{-\alpha}.
\end{equation}
\noindent 
But, in general, the density profile of a cluster
does not have a power-law form,
so $\alpha (\Omega_z,\Lambda_z,M_A,P_k)$
should only be considered as a fitting parameter,
which should, in principle, be
dependent on both the cluster mass and underlying cosmology.
However, motivated by the insight of Navarro, Frenk, \& White (1996)
that there seems to be a universal function
(as a function of scaled radius in units of the virial radius
each individual halo)
for density profiles of dark matter halos,
independent of cosmology and halo mass,
it is hoped that $\alpha$ will only 
be a {\it weak} function of both the underlying
cosmology and cluster mass.
As a matter of fact, as we will show below,
the best fit to N-body results
requires that 
$\alpha$ be a constant equal to $2.3$,
in harmony with the work of Navarro \etal (1996).
Combining equations 8,9,10 yields
\begin{eqnarray}
r_f=\hskip -0.6cm&&3.617 r_A^{(\alpha-3)/\alpha} \left({M_A\over 2.058\times 10^{14}}\right)^{1/\alpha}\nonumber \\ 
\hskip -0.6cm&&\Omega_z^{0.19 (\alpha-3)/\alpha} \Omega_0^{-1/\alpha} C(\Omega_z,\Lambda_z)^{(\alpha-3)/3\alpha}.
\end{eqnarray}
\noindent 
This equation allows us to determine the smoothing
radius $r_f$ for a cluster of mass $M_A$ (within radius $r_A$),
given $\alpha$. 
We see that $r_f$ will be completely deterministic,
{\it if $\alpha$ and $C$ can be specified a priori}.
As we will subsequently show, we will choose to fix $C=1$
and let its dependence on $\Omega_z$ and $\Lambda_z$
be absorbed by another fitting parameter $\delta_c$ (see below).
Therefore, the final equation for $r_f$ that we will use 
is 
\begin{eqnarray}
r_f=\hskip -0.6cm &&3.617 r_A^{(\alpha-3)/\alpha} \left({M_A\over 2.058\times 10^{14}}\right)^{1/\alpha}\nonumber \\
\hskip -0.6cm &&\Omega_z^{0.19 (\alpha-3)/\alpha} \Omega_0^{-1/\alpha}.
\end{eqnarray}
\noindent 
In this equation $\alpha$ is the only adjustable parameter.
But as will be shown later, $\alpha$ turns out to be a constant.
Therefore, 
{\it $r_f$ is no longer an adjustable parameter, rather
it is a unique function of only $M_A$ and $r_A$}.
Another useful expression is to relate $M_v$ to $M_A$ (obtained
by combining equations 9 and 13) in terms of $r_A$:
\begin{eqnarray}
M_v=\hskip -0.6cm &&2.058\times 10^{14} r_A^{3(\alpha-3)/\alpha} \left({M_A\over 2.058\times 10^{14}}\right)^{3/\alpha}\nonumber\\
\hskip -0.6cm && \Omega_z^{0.57 (\alpha-3)/\alpha} \Omega_0^{(\alpha-3)/\alpha} 
\end{eqnarray}
\noindent 
or  
\begin{eqnarray}
M_A=\hskip -0.6cm &&2.058\times 10^{14} r_A^{3-\alpha} \left({M_v\over 2.058\times 10^{14}}\right)^{\alpha/3}\nonumber\\
\hskip -0.6cm &&\Omega_z^{0.19 (3-\alpha)} \Omega_0^{(3-\alpha)/3}.
\end{eqnarray}
\noindent
We note that, in the special case 
where $r_A=r_v$, we have (from equation 9)
\begin{equation}
r_f = \left({M_A\over 4.347\times 10^{12}}\right)^{1/3} \Omega_0^{-1/3},
\end{equation}
\noindent 
which is independent of $\alpha$,
and the virial radius is
\begin{equation}
r_v = \left({M_A\over 2.058\times 10^{14}}\right)^{1/3} \Omega_z^{0.19}\Omega_0^{-1/3}.
\end{equation}
\noindent 
The circular velocity of the (just virialized)
halo at redshift $z$
can be expressed as 
\begin{eqnarray}
v_c\hskip -0.3cm &\equiv&\hskip -0.3cm\left[{GM_v (1+z)\over r_v}\right]^{1/2}\nonumber \\
\hskip -0.3cm&=&\hskip -0.3cm 7.4\times 10^2 \left({M_A\over 1.0\times 10^{14}}\right)^{1/3} \nonumber \\
\hskip -0.3cm&&\hskip -0.3cm\Omega_z^{-0.095}\Omega_0^{1/6}(1+z)^{1/2}~\kms.
\end{eqnarray}
\noindent
Note that the dependences of $v_c$
on $\Omega_z$ and $\Omega_0$ 
are very weak.
But $v_c$ depends rather strongly on $z$ and
somewhat strongly on $M_v$.
The 1-d velocity dispersion $\sigma_{||}$
is just equal to $v_c/\sqrt{2}$.
Another useful relation is between $r_f$ and $v_c$ (or $\sigma_{||}$):
\begin{equation}
r_f =3.845\times 10^{-3} v_c \Omega_z^{0.095}\Omega_0^{-1/2}(1+z)^{-1/2},
\end{equation}
\noindent 
where $v_c$ is in km/s.

Once we have uniquely determined the smoothing window for a given
cluster mass and cosmology (equation 13), 
there is only one last parameter to be specified
before the abundance of peaks forming clusters of a given mass
is uniquely fixed:
we need to specify the required peak height 
such that density peaks with such a height
just collapse and virialize at the redshift in question.
Here once again, we are guided by the spherical perturbation model,
and use $\delta_c$ as a parameter to quantify
the required peak height.
$\delta_c$ is the linear overdensity
of a peak at the concerned redshift
at which a peak just collapses and virializes.
$\delta_c$ is $1.68$ in the spherical top-hat collapse
in an $\Omega_z=1$ model (Gunn \& Gott 1972).
Since realistic density perturbations are likely to be 
non-spherical and non-top-hat
and there are models other than those with $\Omega_z=1$ 
also being considered,
$\delta_c$ can only be treated as a fitting parameter,
to be determined by comparing to N-body simulations.
In other words, the spherical collapse model
is not applied explicitly, rather it is used as a guide
for an initial guess of $\delta_c$.
The actual collapse, or some portion of the collapse in time or in space,
of a density peak may not be spherical.
For example, a triaxial proto-cluster
will require a lower value of $\delta_c$ 
than a spherical proto-cluster (More, Heavens, \& Peacock 1986).
The validity of our method does not critically depend on the
spherical collapse model, rather it is made valid by
comparing to N-body simulations treating $\delta_c$ as a fitting parameter.
Although $\delta_c$ could depend on $P_k$ and cluster virial
masss $M_v$ as well as $\Omega_z$ and $\Lambda_z$,
we will restrict our fitting procedure as if
$\delta_c$ only depends on $\Omega_z$ and $\Lambda_z$.
In any case, the final fit to N-body results
proves that this assumption is good.

We now set down the basic formulae from BBKS
for computing the number density
of peaks of appropriate sizes and heights
for a Gaussian density field.
The differential peak density is
\begin{equation}
N_{pk} (\nu) = {1\over (2\pi)^2 R_*^3} e^{-\nu^2/2} G(\gamma, \gamma\nu),
\end{equation}
\noindent where $G(\gamma,w)$ is, for the convenience of
calculation, 
an analytic formula approximating the exact
three-dimensional integral (see equation A12 of BBKS).
$G(\gamma,w)$ is accurate to better than 1\% over the range
$0.3<\gamma<0.7$ and $-1<w<\infty$
with the accuracy being better than $0.1\%$
for $w>1$, according to BBKS:
\begin{equation}
G(\gamma,w) = {w^3-3\gamma^2 w + [B(\gamma)w^2+C_1(\gamma)]\exp [-A(\gamma)w^2]\over 1+ C_2(\gamma) \exp [-C_3(\gamma) w]}
\end{equation}
\noindent (equation 4.4 of BBKS), where $w\equiv \gamma\nu$.
For all the models which we have computed or are of current interest
we have $\gamma>0.7$ and $w>1.1$ in most cases
with the lowest values being $\gamma=0.58$ and $w=0.84$.
Therefore, it is accurate to use $G(\gamma,w)$ for models
of current interest.
The various symbols are defined as follows.
\begin{eqnarray}
A&=&{5/2\over 9-5\gamma^2} \nonumber\\ 
B&=&{432\over (10\pi)^{1/2}(9-5\gamma^2)^{5/2}} \nonumber\\ 
C_1&=&1.84+1.13(1-\gamma^2)^{5.72} \nonumber\\ 
C_2&=&8.91+1.27 \exp (6.51\gamma^2) \nonumber\\ 
C_3&=&2.58 \exp (1.05\gamma^2) 
\end{eqnarray}
\noindent (equation 4.5 of BBKS).
The parameters $\gamma$ and $R_*$ are related to
the moments of the power spectrum:
\begin{eqnarray}
\gamma&\equiv&{\sigma_1^2\over \sigma_2\sigma_0} \nonumber\\ 
R_*&\equiv&\sqrt{3}{\sigma_1\over\sigma_2}
\end{eqnarray}
\noindent (equation 4.6a of BBKS), and $\sigma_j$ 
are spectral moments:
\begin{equation}
\sigma_j \equiv \int {k^2 dk\over 2\pi^2} P_k k^{2j} W^2(k r_f)
\end{equation}
\noindent (equation 4.6c of BBKS).
We are now ready to derive the cluster mass function 
by counting peaks of appropriate sizes and heights.
The procedure can be described in four steps:

\noindent 1. For a cluster mass $M_A$, we find $r_f$ using equation 13.

\noindent 2. We smooth the power spectrum $P_k$ by the square
of a Gaussian window of radius $r_f$ (equation 6) and 
then compute $\sigma_0$, $\gamma$ and $R_*$ (equations 23,24).

\noindent 3. Requiring that $\nu \sigma_0 = \delta_c$ yields the threshold
peak height $\nu_t=\delta_c/\sigma_0$.

\noindent 4. Integrating equation 20 from $\nu_t$ to $\infty$
gives the cumulative cluster mass function:
\begin{equation}
n(>M_A)=\int_{\nu_t}^\infty N_{pk} (\nu) d\nu.
\end{equation}
\noindent 

To summarize, we have three adjustable parameters:
$\alpha (\Omega_z,\Lambda_z,M_A,P_k)$, $\delta_c (\Omega_z,\Lambda_z)$
and $C(\Omega_z,\Lambda_z)$.
In general, 
$\alpha(\Omega_z,\Lambda_z,M_A,P_k)$ should be a function 
of four variables, $\Omega_z$, $\Lambda_z$, $M_A$ and $P_k$,
whereas $\delta_c$ and $C$
should depend only on $\Omega_z$ and $\Lambda_z$.
We choose to fix $C(\Omega_z,\Lambda_z)$ to be unity, 
independent of cosmological parameters, and
treat only $\alpha$ and $\delta_c$ as two adjustable parameters.
Since both parameters 
($C$ and $\delta_c$)
depend on the same cosmological parameters,
this treatment is justified and simplifies 
the fitting procedure;
the dependence of $C(\Omega_z,\Lambda_z)$ on cosmological
parameters is assumed to be absorbed by $\delta_c(\Omega_z,\Lambda_z)$.

\begin{deluxetable}{cccccccc} 
\tablewidth{0pt}
\tablenum{1}
\tablecolumns{6}
\tablecaption{List of parameters for 32 models} 
\tablehead{
\colhead{Model} &
\colhead{$\Omega_0$} &
\colhead{$\Lambda_0$} &
\colhead{$\sigma_8$} &
\colhead{$EP\equiv 3.4\sigma_{25}/\sigma_8$} &
\colhead{Comment}}

\startdata
$1$ & $1.000$ & $0.000$ & $1.050$ & 0.956 & SCDM$\tablenotemark{a}$\nl  
$2$ & $1.000$ & $0.000$ & $0.700$ & 0.956 & SCDM$\tablenotemark{a}$\nl  
$3$ & $1.000$ & $0.000$ & $0.525$ & 0.956 & SCDM$\tablenotemark{a}$\nl  
$4$ & $1.000$ & $0.000$ & $0.350$ & 0.956 & SCDM$\tablenotemark{a}$\nl  
$5$ & $1.000$ & $0.000$ & $1.050$ & 1.088 & $P_k=k^{-1}$\nl  
$6$ & $1.000$ & $0.000$ & $0.808$ & 1.088 & $P_k=k^{-1}$\nl  
$7$ & $1.000$ & $0.000$ & $0.700$ & 1.088 & $P_k=k^{-1}$\nl  
$8$ & $1.000$ & $0.000$ & $0.525$ & 1.088 & $P_k=k^{-1}$\nl  
$9$ & $1.000$ & $0.000$ & $0.350$ & 1.088 & $P_k=k^{-1}$\nl  
$10$ & $1.000$ & $0.000$ & $1.050$ & 1.923 & $P_k=k^{-2}$\nl  
$11$ & $1.000$ & $0.000$ & $0.808$ & 1.923 & $P_k=k^{-2}$\nl  
$12$ & $1.000$ & $0.000$ & $0.700$ & 1.923 & $P_k=k^{-2}$\nl  
$13$ & $1.000$ & $0.000$ & $0.525$ & 1.923 & $P_k=k^{-2}$\nl  
$14$ & $0.350$ & $0.000$ & $0.800$ & 1.196 & OCDM1$\tablenotemark{b}$\nl    
$15$ & $0.410$ & $0.000$ & $0.689$ & 1.196 & OCDM1$\tablenotemark{b}b$\nl    
$16$ & $0.446$ & $0.000$ & $0.632$ & 1.196 & OCDM1$\tablenotemark{b}$\nl 
$17$ & $0.520$ & $0.000$ & $0.524$ & 1.196 & OCDM1$\tablenotemark{b}$\nl 
$18$ & $0.600$ & $0.000$ & $1.000$ & 1.060 & OCDM2$\tablenotemark{b}$\nl 
$19$ & $0.661$ & $0.000$ & $0.818$ & 1.060 & OCDM2$\tablenotemark{b}$\nl 
$20$ & $0.692$ & $0.000$ & $0.730$ & 1.060 & OCDM2$\tablenotemark{c}$\nl 
$21$ & $0.750$ & $0.000$ & $0.575$ & 1.060 & OCDM2$\tablenotemark{c}$\nl 
$22$ & $0.818$ & $0.000$ & $0.404$ & 1.060 & OCDM2$\tablenotemark{c}$\nl 
$23$ & $0.400$ & $0.600$ & $0.790$ & 1.225 & LCDM1$\tablenotemark{d}$\nl 
$24$ & $0.692$ & $0.308$ & $0.591$ & 1.225 & LCDM1$\tablenotemark{d}$\nl 
$25$ & $0.842$ & $0.158$ & $0.460$ & 1.225 & LCDM1$\tablenotemark{d}$\nl 
$26$ & $0.200$ & $0.800$ & $1.500$ & 1.225 & LCDM2$\tablenotemark{e}$\nl 
$27$ & $0.355$ & $0.645$ & $1.321$ & 1.225 & LCDM2$\tablenotemark{e}$\nl 
$28$ & $0.458$ & $0.542$ & $1.211$ & 1.225 & LCDM2$\tablenotemark{e}$\nl 
$29$ & $0.573$ & $0.427$ & $1.088$ & 1.225 & LCDM2$\tablenotemark{e}$\nl 
$30$ & $0.667$ & $0.333$ & $0.982$ & 1.225 & LCDM2$\tablenotemark{e}$\nl 
$31$ & $0.796$ & $0.204$ & $0.813$ & 1.225 & LCDM2$\tablenotemark{e}$\nl 
$32$ & $0.871$ & $0.129$ & $0.690$ & 1.225 & LCDM2$\tablenotemark{e}$\nl 
\enddata

\tablenotetext{a}{the standard CDM model with Hubble constant
$H_o=50$km/s/Mpc, $\Omega_0=1.0$ and $n=1.0$,
where $n$ is the power index at very large scale.
BBKS power spectrum transfer function (equation G3) is used}

\tablenotetext{b}{an open CDM model with Hubble constant
$H_o=70$km/s/Mpc, $\Omega_0=0.35$ and $n=1.0$;
BBKS power spectrum transfer function (equation G3) is used.}

\tablenotetext{c}{an open CDM model with Hubble constant
$H_o=60$km/s/Mpc, $\Omega_0=0.60$ and $n=1.0$;
BBKS power spectrum transfer function (equation G3) is used.}

\tablenotetext{d}{a CDM model with a cosmological
constant with Hubble constant
$H_o=65$km/s/Mpc, $\Omega_0=0.40$, $\Lambda_0=0.60$
and $n=0.95$;
the power spectrum transfer function is computed as in Cen \etal 1993.}

\tablenotetext{e}{a CDM model with a cosmological
constant with Hubble constant
$H_o=100$km/s/Mpc, $\Omega_0=0.20$, $\Lambda_0=0.80$ and $n=0.95$;
the same power spectrum as LCDM1 is used.}
\end{deluxetable}

We now turn to N-body simulations to calibrate 
GPM to compute the cluster mass function,
i.e., to determine the two fitting 
parameters in GPM, $\alpha$ and $\delta_c$.
We have thirty two cosmological models at our disposal
to calibrate GPM and test its accuracy.
The models are listed in Table 1.
The second and third columns are the density parameter
and cosmological constant of the model, respectively.
The fourth column is the linear rms density fluctuation
in an $8~h^{-1}$Mpc top-hat sphere at $z=0$.
The fifth column, $EP$ (Excess Power),
is a parameter to 
describe the shape of the power spectrum 
on scales $\sim 8-300\;h^{-1}$Mpc,
introduced by Wright \etal (1992).
The definition is $EP\equiv 3.4\sigma_{25}/\sigma_8$,
where $\sigma_{25}$ is the linear rms density fluctuation 
in a $25\;h^{-1}$Mpc top-hat sphere at $z=0$.
The last column indicates the type of
power spectrum used (details are given in the table footnotes).

Some models are physically
self-consistent in the sense that
their power spectrum transfer functions
are computed for the given cosmological parameters,
while others are not.
The latter are included
to increase the coverage of the parameter
space for calibration purposes.
Taken together these thirty two models span 
the ranges of $\Omega_z$, $\Lambda_z$, $\sigma_8$ and $EP$ of current 
interest:
$\Omega_z=0.2\rightarrow 1.0$,
$\Lambda_z=0.0\rightarrow 0.8$,
$\sigma_8=0.35\rightarrow 1.5$,
and $EP=0.735\rightarrow 1.923$.
As a note, Wright \etal (1992) find that 
the range of $EP$ that fits the COBE data 
is $1.30\pm 0.15$ ($1\sigma$),
which is consistent with analysis of the
galaxy power spectrum by 
Peacock \& Dodds (1994) and Feldman, Kaiser \& Peacock (1994),
or observations of large-scale galaxy clustering by
Maddox \etal (1990).
The reason that we use $EP$ rather than $\Gamma$ ($=\Omega_0 h$)
is that $\Gamma$ can only be used for CDM
type models with $n=1$.

Each model is run using 
a particle-mesh code with a box size $400\;h^{-1}$Mpc.
A large simulation box is needed in
order to produce a significant
number of the rich but rare clusters.
The simulation
box contains $720^{3}$ cells and $240^{3} = 10^{7.1}$ dark matter
particles, with a particle mass of $1.3 \times 10^{12} \Omega_0\;h^{-1}\msun$. 
The mass resolution seems adequate for our purpose:
a cluster of mass $6\times 10^{14}h^{-1}\msun$ contains
$462\Omega_{0}^{-1}$ particles.
In each simulation, 
clusters are selected as the maxima of the mass distribution  within
spheres of comoving radius of $r_A=1.5\rm h^{-1}$ Mpc.
The mass of each cluster is determined in a sphere within 
the Abell radius $r_A$.
The results are not sensitive to the cluster-finding algorithm
{\it as long as the mass is defined within a chosen radius}.
In other words, different group-finding algorithms such as friends-of-friends
or DENMAX are all able to locate the density peaks properly.

\begin{figure*}
\centering
\begin{picture}(400,300)
\psfig{figure=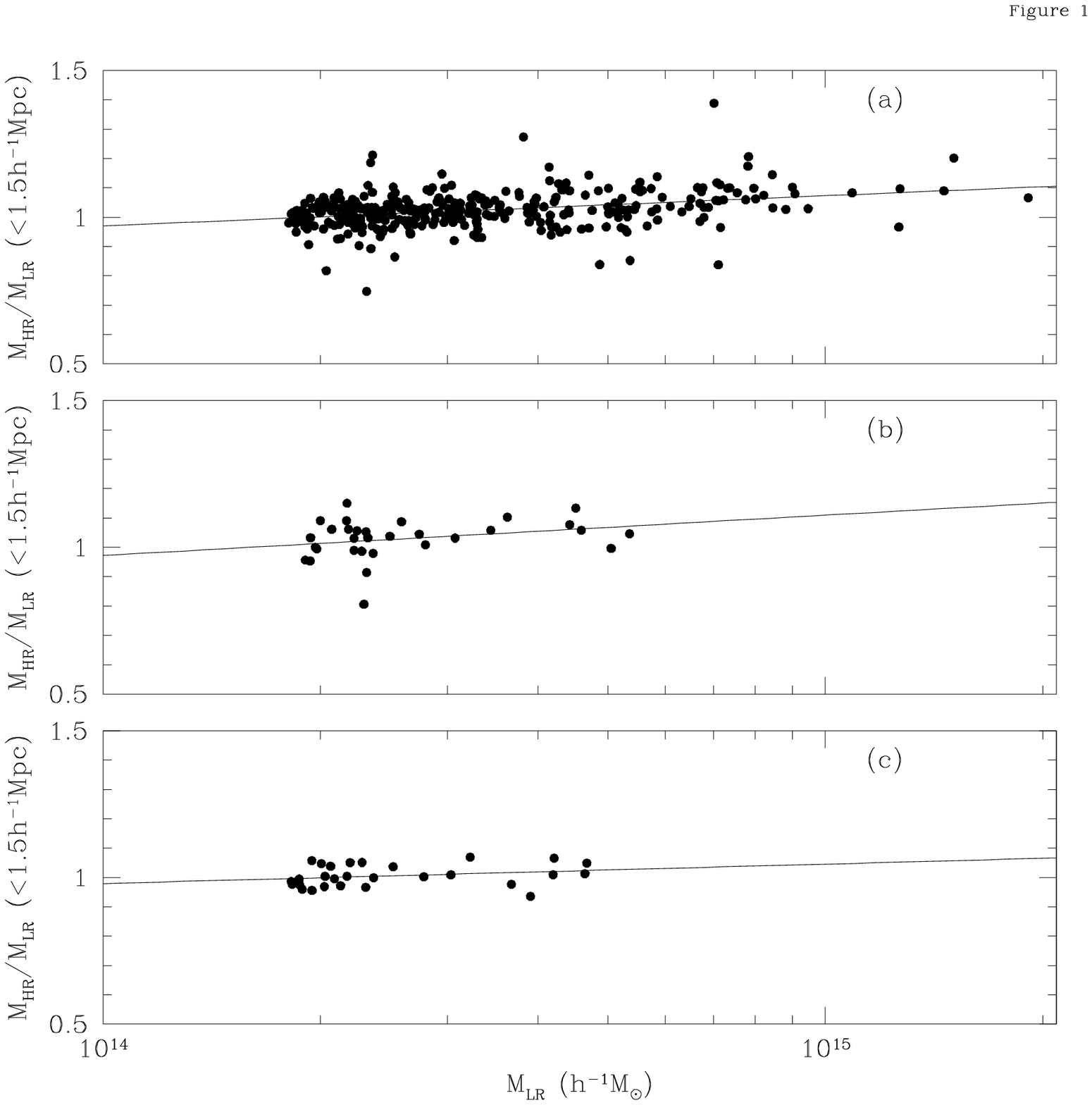,height=10.0cm,width=15.0cm,angle=0.0}
\end{picture}
\caption{
$M_{HR}/M_{LR}$ is plotted against $M_{LR}$ for
the three models.
The solid line in each plot is the best linear fit:
$M_{HR}/M_{LR}=a + b \log_{10}M_{LR}$, where
$M_{LR}$ is in $h^{-1}\msun$.
We find $(a,b)$ to be
$(-0.469,0.103)$,
$(-0.0528,0.0662)$,
$(-0.952,0.137)$ for models (1,14,29), respectively.}
\end{figure*}

\begin{figure*}
\centering
\begin{picture}(400,300)
\psfig{figure=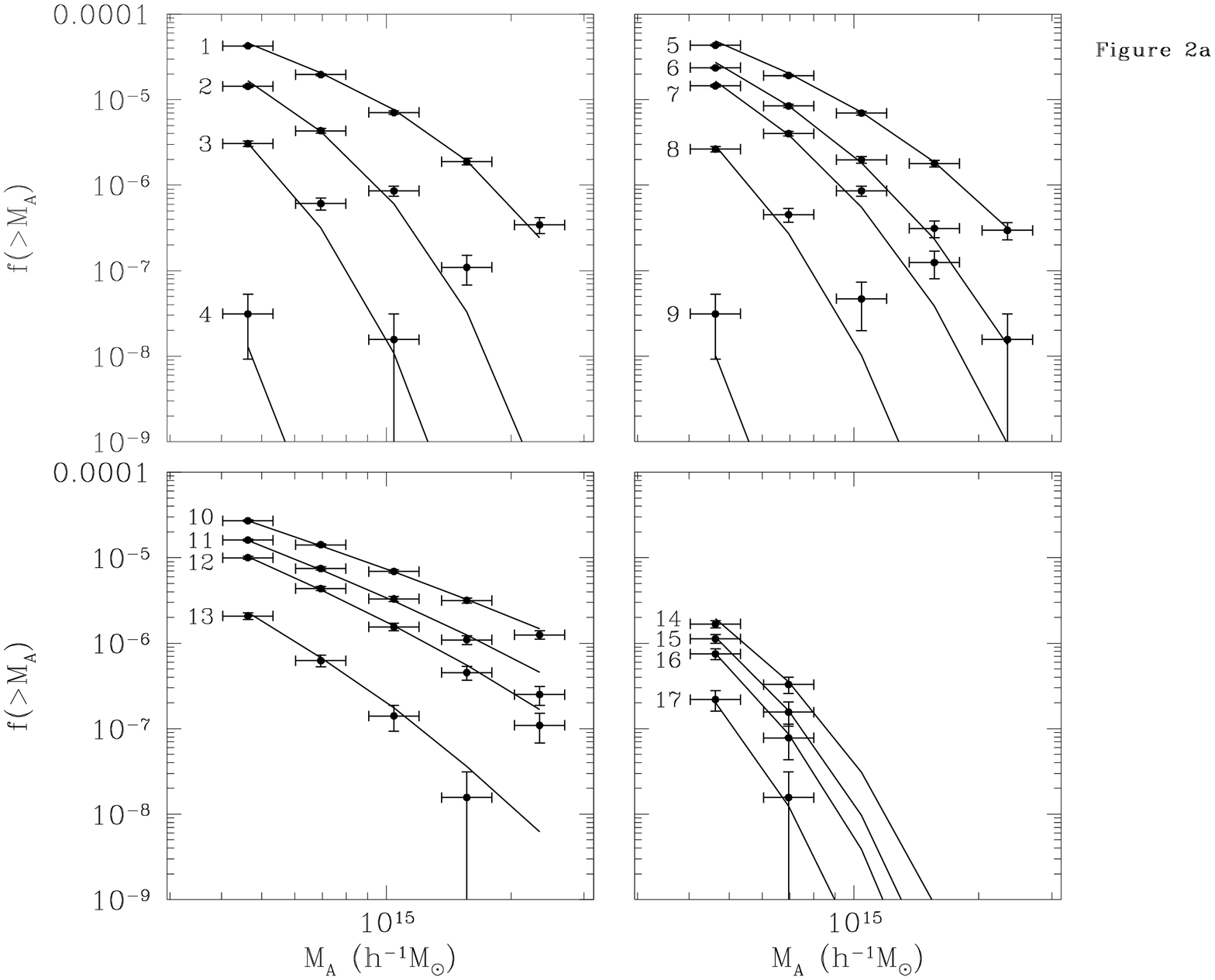,height=10.0cm,width=15.0cm,angle=-90.0}
\end{picture}
\centerline{(2a)}\vspace{0.1in}
\end{figure*}

\begin{figure*}
\centering
\begin{picture}(400,260)
\psfig{figure=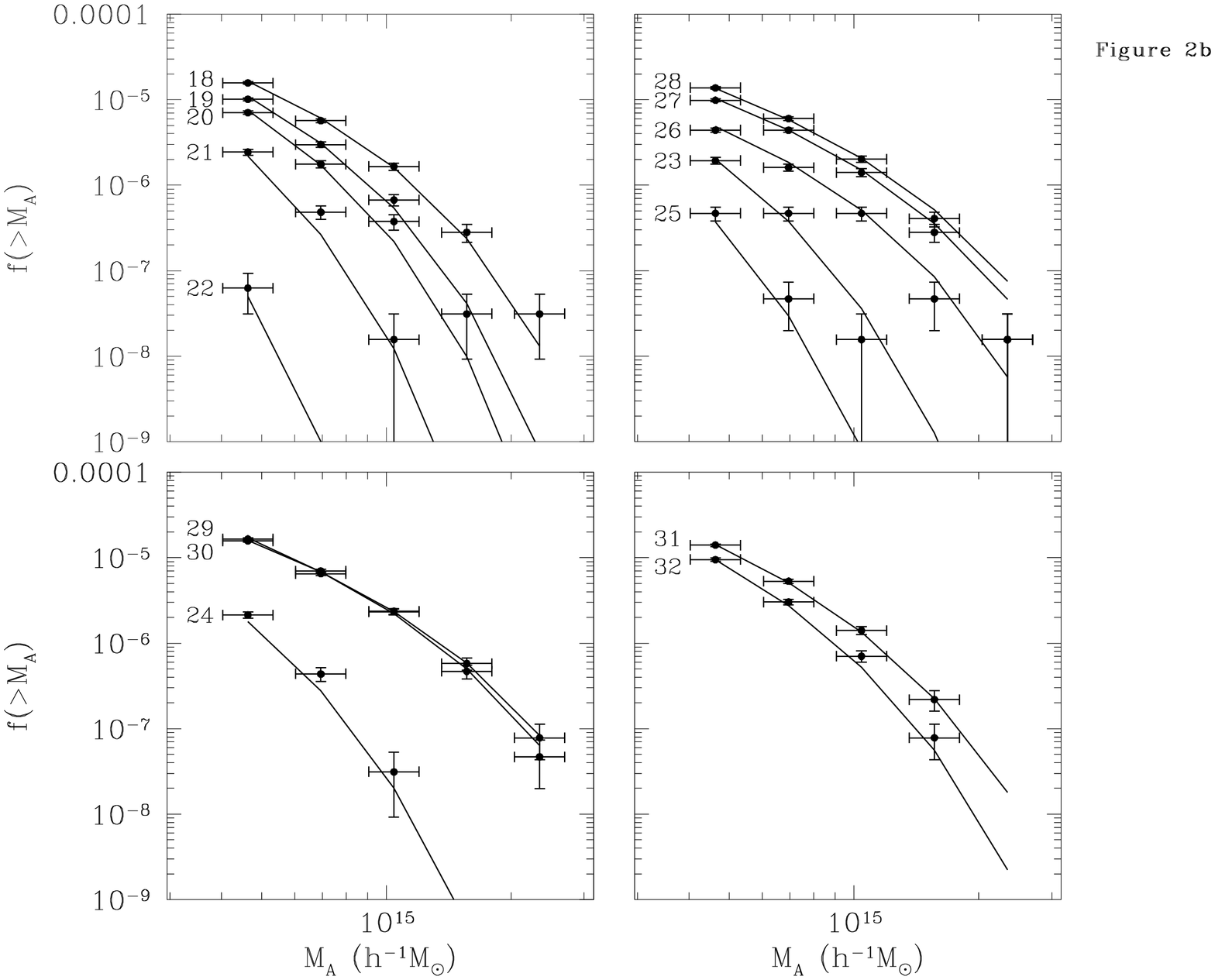,height=10.0cm,width=15.0cm,angle=-90.0}
\end{picture}
\centerline{(2b)}\vspace{0.1in}
\caption{
Mass functions of various models.
Each curve is labeled by its model number from Table 1.
The simulation results are shown
as symbols with horizontal errorbars
being the uncertainties in the mass determination (15\%)
and the vertical errorbars being the statistical $1\sigma$ errorbars
for the number of clusters.
The solid curves in Figure 2
are the results from GPM.
Note that simulation box size limits the density
of clusters to $>1.56\times 10^{-8}h^3$Mpc$^{-3}$, when there is only
one cluster in the whole box.
At $f(>M_A)=10^{-6}h^3$Mpc$^{-3}$, there are 65 clusters in the simulation
box.
}
\end{figure*}

Before we present our numerical results, 
it is important to check resolution effect.
We make the following resolution calibration test.
We run two simulations for the SCDM model (model 1 
in Table 1)
with an identical initial realization in a box of size $L=128h^{-1}$Mpc.
One of the two simulations uses
the same PM code as for the thirty two models listed in Table 1
and has the same numerical resolution ($1.39\;h^{-1}$Mpc); 
the other simulation has a much higher resolution ($0.0625\;h^{-1}$Mpc)
based on the P$^3$M scheme utilizing a special computer
chip (GRAPE) to solve the PP part of the force computation
(Brieu, Summers \& Ostriker 1995).
Since the resolution element of the P$^3$M simulation 
is much smaller than the Abell radius,
it can be considered as having an infinite resolution 
(i.e., representing the truth) for 
the purpose of calibrating our low resolution results.
Because the two simulations have identical initial conditions,
we are able to identify every rich cluster 
in one simulation with its counterpart in the other.
Having made such a one-to-one correspondence
we can compute the ratio $M_{HR}/M_{LR}$
($HR$ stands for high resolution and $LR$ for low resolution)
as a function of cluster mass $M_{LR}$.
This allows us to make corrections to $M_{LH}$ 
in the lower resolution simulation.
The above resolution calibration procedure is repeated for  
an open CDM model (model 14 in Table 1)
and a CDM model with a cosmological constant
(model 23 in Table 1).

Figures (1a,b,c) show
the results for the three models,
where $M_{HR}/M_{LR}$ is plotted against $M_{LR}$.
The solid line in each plot is the best linear fit:
$M_{HR}/M_{LR}=a + b \log_{10}M_{LR}$, where
$M_{LR}$ is in $h^{-1}\msun$.
We find $(a,b)$ to be 
$(-0.469,0.103)$,
$(-0.0528,0.0662)$,
$(-0.952,0.137)$ for models (1,14,23), respectively.
Let us call the three fitting functions ($M_{HR}/M_{LR}$)
as $R_1(M)$, $R_{035}(M)$, $R_{04}(M)$,
for the three models run:
$\Omega_z=1$ (model 1 in Table 1),
$\Omega_z=0.35$ and $\Lambda_z=0$ (model 14), and 
$\Omega_z=0.40$ and $\Lambda_z=0.60$ (model 23).
Then, for the mass of each cluster, $M_{LR}$, 
we correct it by multiplying it
by 
$R_1 + (R_{035}-R_1) (1-\Omega_z)/0.65$ in an open model,
or by
$R_1 + (R_{04}-R_1) (1-\Omega_z)/0.60$ in a $\Lambda$ model,
where $\Omega_z$ is the density parameter of the model
under consideration.
From Figure 1
we see that the typical correction is about 5-10\% in the upward direction
with a dispersion of $\sim 5\%$;
i.e., lower resolution simulations slightly underestimate
the mass within the Abell radius, as expected.
Calibrating the lower resolution simulation results
we assign an errorbar of $15\%$ for each cluster mass.

We are now ready to find the best fitting
parameters ($\alpha$, $\delta_c$) by
comparing results from GPM to the direct N-body results.
Before starting the fitting procedure, we have 
some rough idea about what the values of $\alpha$ and $\delta_c$
may be. We pick $\alpha=2.5$ and $\delta_c=1.5$ as an initial guess.  
In the end, the best values are found to be
\begin{equation}
\alpha=2.3 ,
\end{equation}
\noindent 
a constant independent of the cluster mass and cosmology, and
\begin{equation}
{\delta_c=\cases{&\hskip -0.5cm$1.40-0.01(1.0-\Omega_z)\quad\quad\hbox{for}\quad \Lambda_z=0$ \cr 
&\hskip -0.5cm$1.40+0.10(1.0-\Omega_z)\quad\quad \hbox{for}\quad \Omega_z+\Lambda_z=1$.\cr}}
\end{equation}
\noindent 
The best overall fit for all the models
is judged by the author by direct visual examination.
We find it very difficult to design an automated
fitting procedure to be gauged by some objective parameters,
because of the enormous range of the number densities of 
clusters and hard-to-define errorbars hence weighting 
schemes for the densities.
But as we will see, the final fits are probably 
as good as one would have hoped,
which suggests that our somewhat subjective
fitting procedure works very well.
In any case,
the final fit values span very narrow ranges (in fact, $\alpha$
turns out to a constant,
and 
$\delta_c$ varies from 1.40 to 1.39
from $\Omega_z=1$ to $0$ in $\Lambda_z=0$ models,
from 1.40 to 1.36
from $\Omega_z=1$ to $0$ in $\Lambda_z+\Lambda_z=1$ models),
which indicates that the fitting procedure is robust and stable.
We note that the fitted value of $\alpha$ is in excellent agreement
with observations (Carlberg \etal 1996; Fischer \etal 1997).
It seems useful to estimate the uncertainties
in the fitted values of $\alpha$ and $\delta_c$.
However, the sensitivity of a fit to the two parameters
depends on the fitted mass function itself:
a low amplitude mass function depends more sensitively on the two parameters
than a high amplitude mass function.
This is so, of course, because the abundance of rarer
objects depends more sensitively on the parameters.
Roughly speaking, $\alpha$ serves more to fix the shape of the
mass function in a somewhat less sensitive way,
while $\delta_c$ determines the overall amplitude and more sensitively
the amplitude on the high mass end of the mass function.
Our estimates on the uncertainties
are $\Delta \alpha=0.1$ and $\Delta\delta_c=0.01$.

Figures (2a,b) show
the simulation results as symbols 
for 32 models with horizontal errorbars
being the uncertainties in the mass determination (15\%)
and the vertical errorbars being the statistical $1\sigma$ errorbars.
Note that simulation box size limits the density
of clusters to $>1.56\times 10^{-8}h^3$Mpc$^{-3}$, when there is only
one cluster in the whole box.
At $f(>M_A)=10^{-6}h^3$Mpc$^{-3}$, there are 65 clusters.
The solid curves in Figure 2 
are the results from GPM.
We see that the GPM results fit remarkably
well the simulation results for all
the thirty two models examined.
We note that the actual errorbars should be larger
than what are shown for the simulated results
because of cosmic variances;
i.e., the simulation boxsize, although quite large being $400\;h^{-1}$Mpc,
may still not be large enough to have the cosmic variance diminished,
especially for models
with significant power on several
hundred megaparsecs scales.
In any case, the GPM results
fit the N-body results for all the models 
within $2\sigma$ in the vertical axis,
and within a factor of $1.25$ in the horizontal axis.
Since the observed mass function (BC)
has uncertainties in mass about a factor $2.0$ 
and in number density about a factor $\ge 2.0$,
the GPM results are practically 
precise,
for the purpose of comparing model results computed using
GPM with observations.

At this point it seems appropriate to 
reiterate the virtue of the current method.
The essential unique ingredients
are the introduction of two adjustable parameters,
$\alpha$ and $\delta_c$,
the first 
of which turns out to be a constant and
the second of which can be simply expressed
as a function of $\Omega_z$ (the density parameter at
the redshift in question).
Note that the fitted parameter $\delta_c(\Omega_z)$
has a slightly different form for the case with $\Lambda_z=0$
than for the case with $\Omega_z+\Lambda_z=1$.
The fact that $\delta_c$ is only a rather weak function
of $\Omega_z$ in both cases
indicates that the method is robust.

\subsection{Gaussian Peak Method for Cluster Correlation Function}

Having found that the initial density peaks, appropriately defined,
indeed correspond to the clusters formed at late times,
as indicated by the goodness of the fits
of the results from GPM to the N-body results
in terms of cluster mass function presented in the preceding section,
we have some confidence that 
we may be able to compute the cluster-cluster two-point
correlation function using GPM.
We will now proceed along this route.

\begin{figure*}
\centering
\begin{picture}(400,300)
\psfig{figure=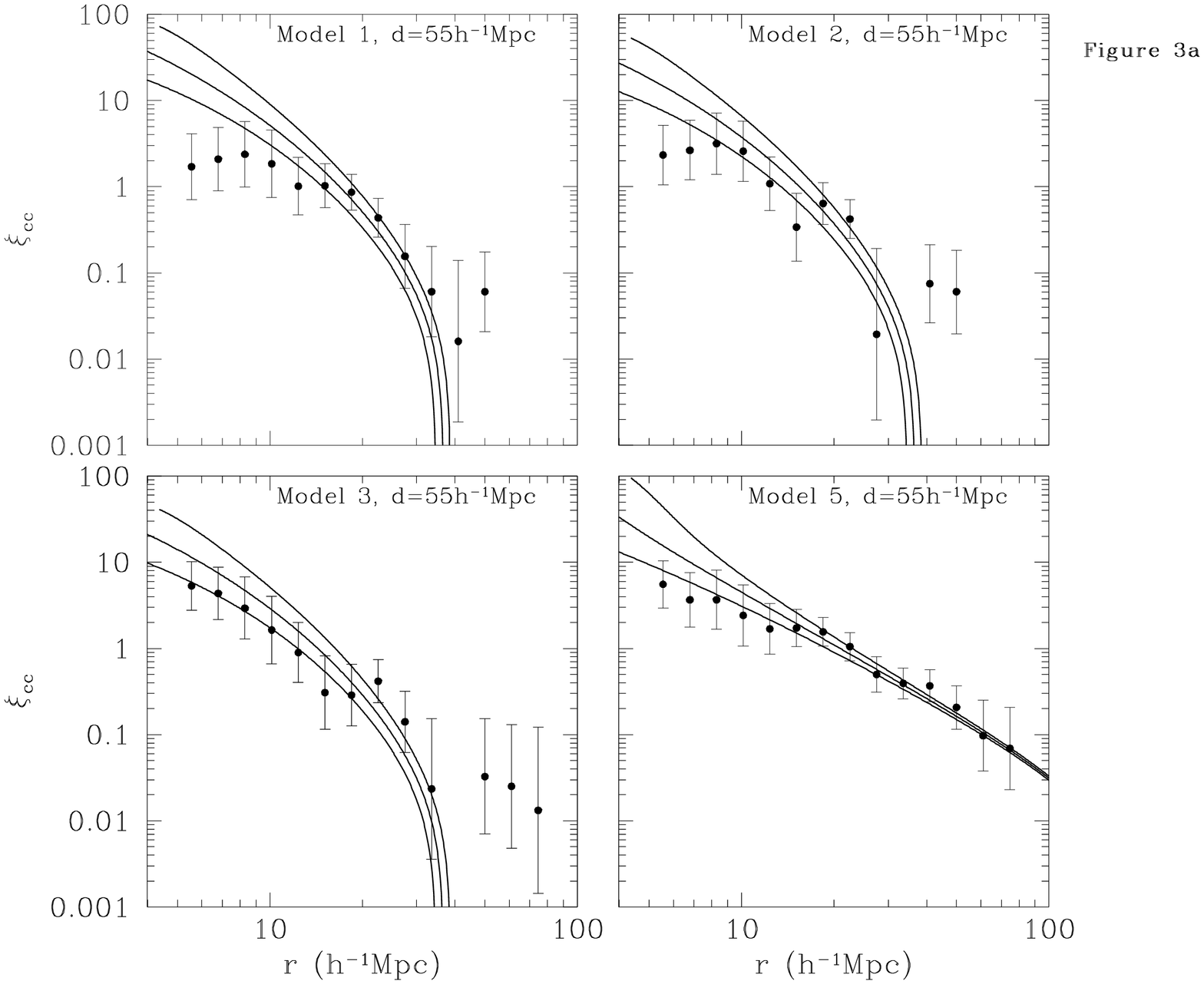,height=10.0cm,width=15.0cm,angle=-90.0}
\end{picture}
\centerline{(3a)}\vspace{0.1in}
\end{figure*}

\begin{figure*}
\centering
\begin{picture}(400,300)
\psfig{figure=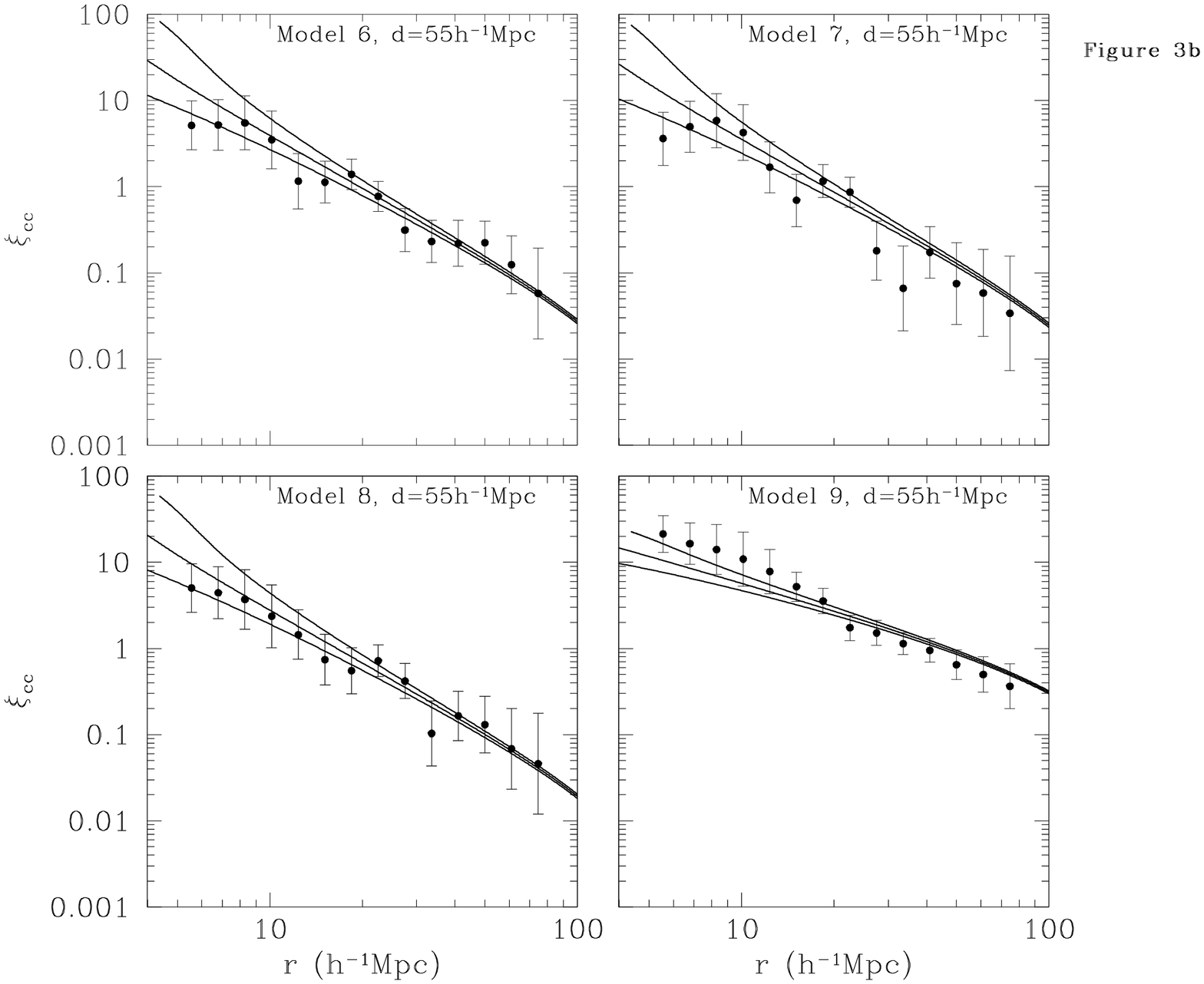,height=10.0cm,width=15.0cm,angle=-90.0}
\end{picture}
\centerline{(3b)}\vspace{0.1in}
\end{figure*}

\begin{figure*}
\centering
\begin{picture}(400,300)
\psfig{figure=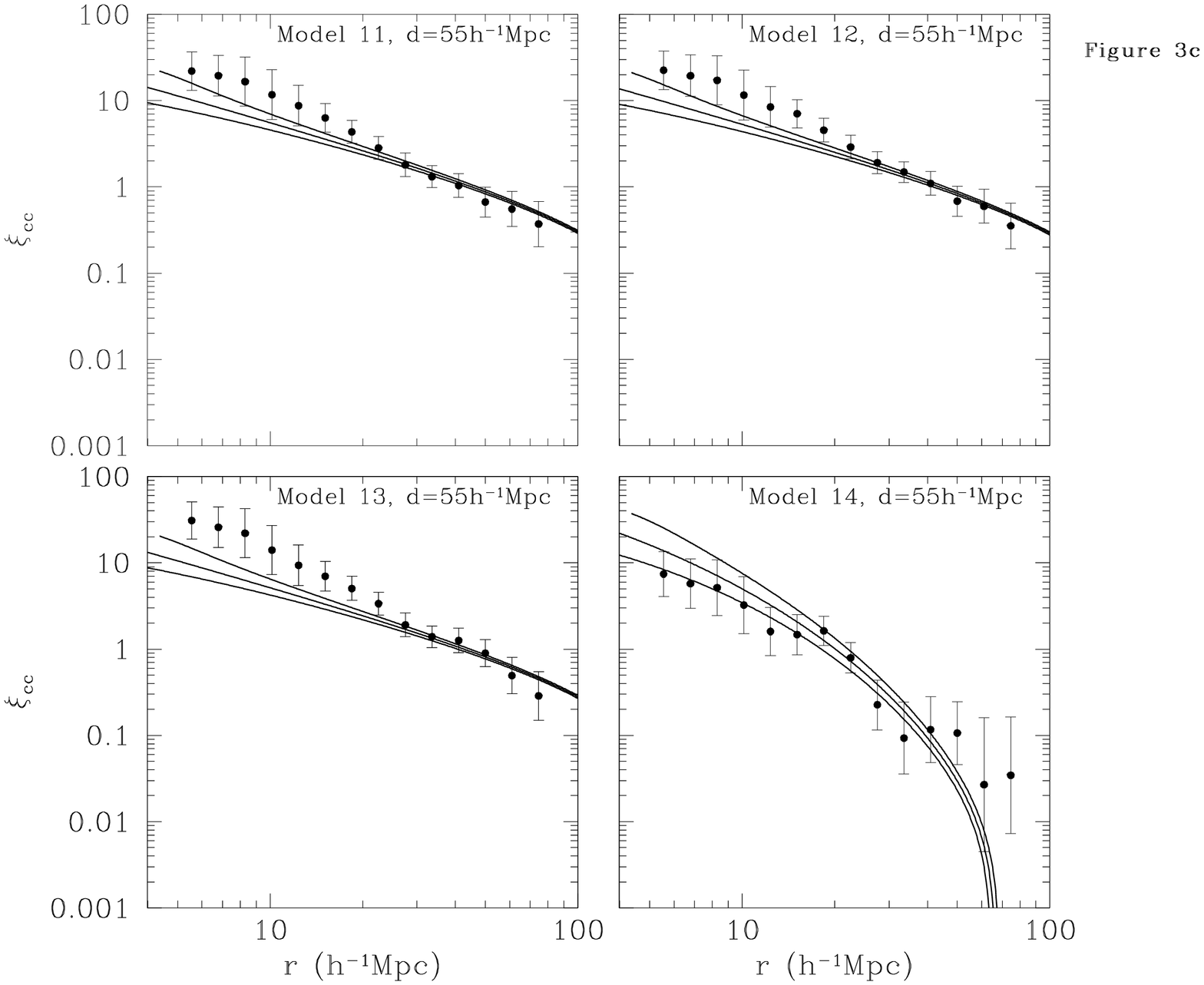,height=10.0cm,width=15.0cm,angle=-90.0}
\end{picture}
\centerline{(3c)}\vspace{0.1in}
\end{figure*}

\begin{figure*}
\centering
\begin{picture}(400,300)
\psfig{figure=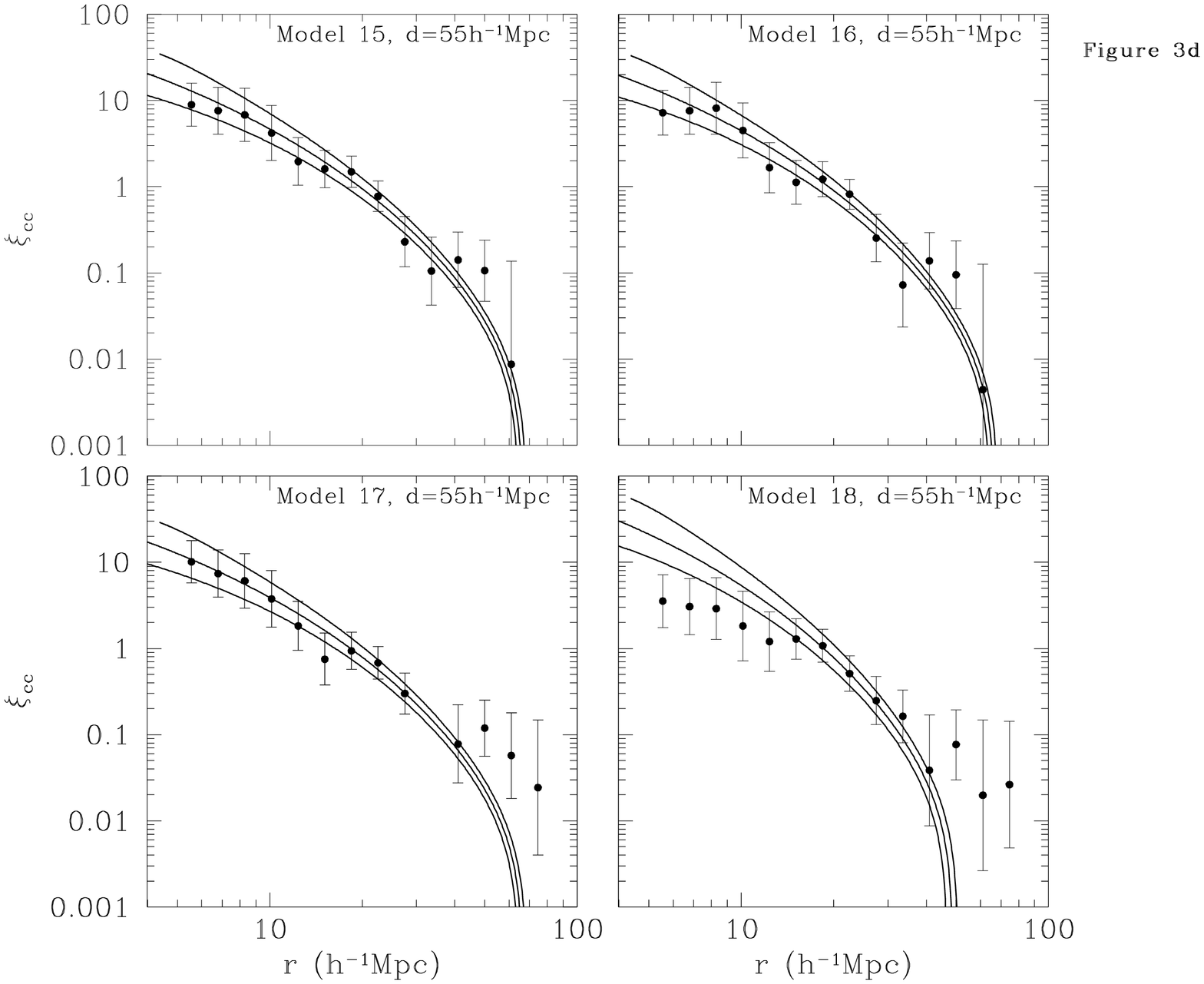,height=10.0cm,width=15.0cm,angle=-90.0}
\end{picture}
\centerline{(3d)}\vspace{0.1in}
\end{figure*}

\begin{figure*}
\centering
\begin{picture}(400,300)
\psfig{figure=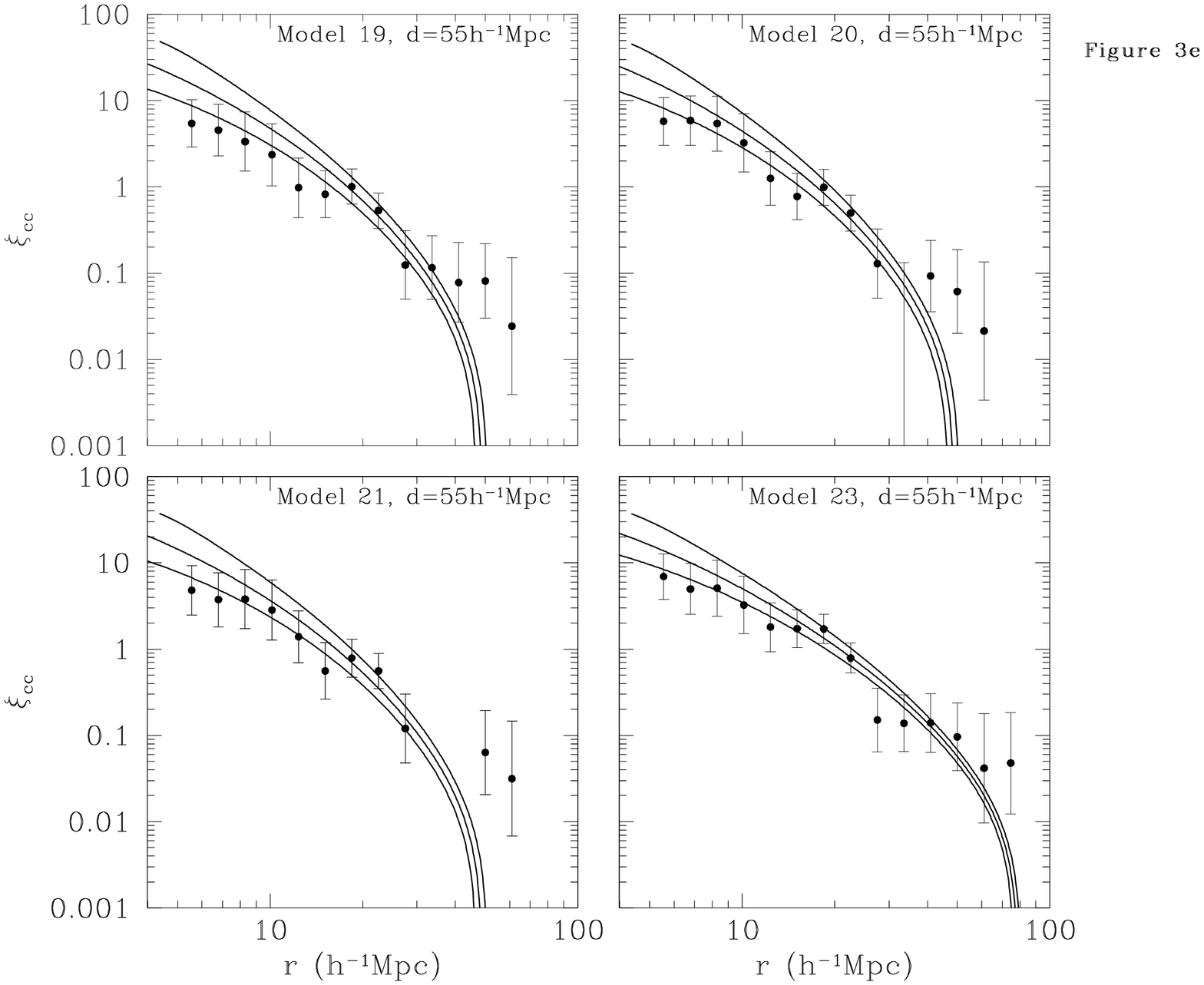,height=10.0cm,width=15.0cm,angle=-90.0}
\end{picture}
\centerline{(3e)}\vspace{0.1in}
\end{figure*}

\begin{figure*}
\centering
\begin{picture}(400,300)
\psfig{figure=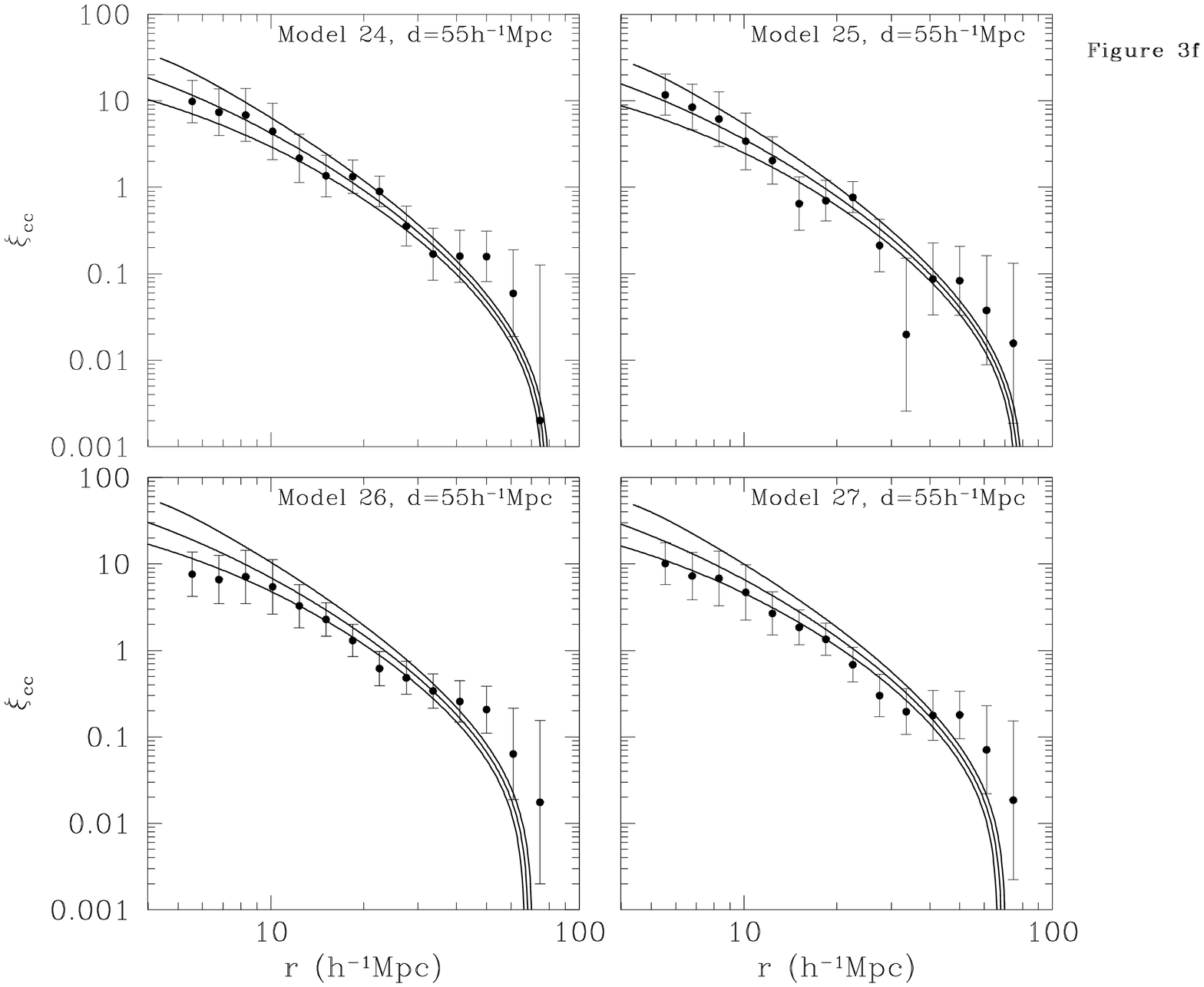,height=10.0cm,width=15.0cm,angle=-90.0}
\end{picture}
\centerline{(3f)}\vspace{0.1in}
\end{figure*}

\begin{figure*}
\centering
\begin{picture}(400,270)
\psfig{figure=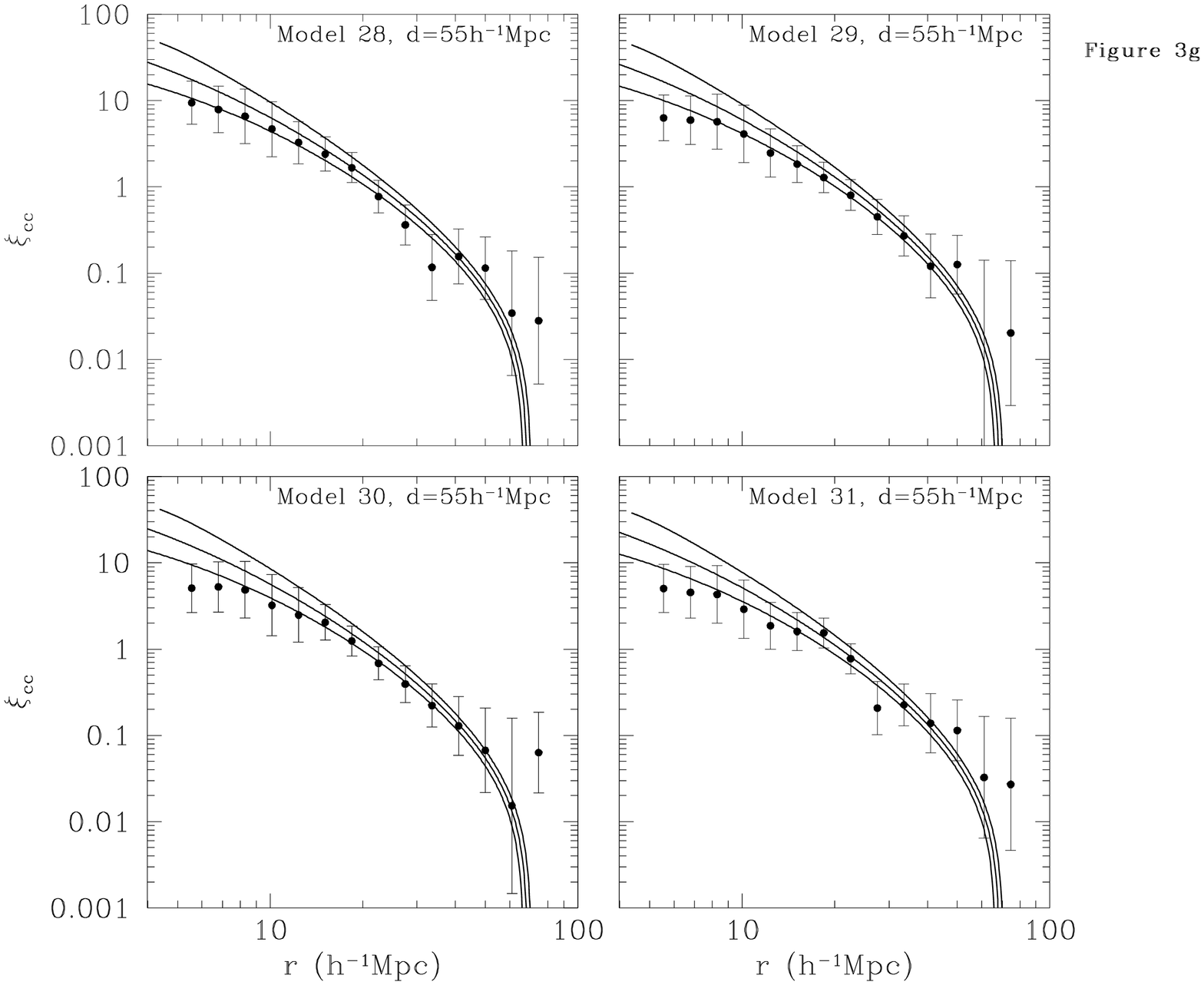,height=10.0cm,width=15.0cm,angle=-90.0}
\end{picture}
\centerline{(3g)}\vspace{0.1in}
\caption{
The cluster correlation functions
for clusters with mean separation of $55h^{-1}$Mpc,
for 28 models as indicated in the panels.
The errorbars are $1\sigma$ statistical.
Three curves are shown in each panel for each model;
the middle curve is obtained using equation 28
and the top and bottom curves are obtained by
adding $\pm 2h^{-1}$Mpc to each point
of the middle curve in the x-axis.
}
\end{figure*}

Even in the linear regime where dynamic contribution 
to the clustering can be ignored,
the primary difficulty in calculating the cluster-cluster two-point
correlation function using Gaussian peaks 
is the ambiguity of relating appropriate
peaks to the clusters of interest.
This ambiguity is reflected in one's
inability to fix the smoothing length $r_f$ and
the threshold peak height $\nu_t$.
The GPM, described in \S 2, eliminates this ambiguity by
simultaneously fixing both $r_f$ and $\nu_t$.
This is achieved by 
demanding that the appropriate peaks
yield the correct cluster mass function,
when compared to direct N-body simulations.

Following BBKS, 
we use the following approximate formula, which is applicable 
when the correlation function is smaller than unity
[however, BBKS state that it may well be a reasonable 
approximation even when the statistical correlation function 
(first term at the right hand side of equation 28, see below)
is not really small],
to compute the final cluster-cluster
correlation function including linear dynamical 
contributions:
\begin{equation}
\xi_{pk,pk}\approx \left({<\tilde\nu>\over \sigma_0} + 1\right)^2\xi_{\rho,\rho},
\end{equation}
\noindent 
(equation 6.63 of BBKS),
where $\xi_{\rho,\rho}$ is the two-point density auto-correlation 
function, and $<\tilde\nu>$ is defined by
\begin{equation}
\tilde\nu\equiv\int_{\nu_t}^\infty [\nu-\gamma\theta/(1-\gamma^2)] N_{pk} d\nu
\end{equation}
\noindent 
(equation 6.45 of BBKS),
where $\nu_t$ is determined in the procedure
described in \S 2; $\gamma$ is defined in equation 23;
$N_{pk}$ is defined in equation 20.
Note that we have chosen a step function for threshold function 
$t(\nu/\nu_t)$
in equation 6.45 of BBKS to arrive at equation 29, i.e.,
only peaks above $\nu_t$ are assumed to be able to collapse
and no peaks below $\nu_t$ are allowed to collapse.
A smoother threshold function (e.g., Kaiser \& Davis 1985; BBKS)
may be used, but the primary effect will be to slightly change the fitting
parameter $\delta_c$.
We therefore will use the sharp step function as the threshold
function without loss of generality.
$\theta$ is defined by
\begin{equation}
\theta\equiv{3(1-\gamma^2)+(1.216-0.9\gamma^4) \exp [-\gamma/2(\gamma\nu/2)^2]\over [3(1-\gamma^2) + 0.45 + (\gamma\nu/2)^2]^{1/2} + \gamma\nu/2}
\end{equation}
\noindent 
(equation 6.14 of BBKS).
We note that equation 28 is 
valid in the linear regime when
$\xi_{\rho,\rho}$ is much less than unity and
it is not yet clear whether the approximation, coupled
with our definitive peak identification method, 
also works in the mild nonlinear regime.
Our goal is to find an approximation based on equation 28 which
will give sufficiently accurate results 
for $\xi_{pk,pk}$ in the regime
whose values are of order unity and below.
For this reason we choose to modify equation 28
in the following manner:
$\xi_{pk,pk}\approx \left({<\tilde\nu>\over \sigma_0} + 1\right)^2\xi_{\rho,\rho}D(\bar\xi_{\rho,\rho})$,
where $D(\bar\xi_{\rho,\rho})$ is a fitting parameter that
depends only on $\bar\xi_{\rho,\rho}$, which is the mean
two-point matter correlation function, defined as
$\bar\xi_{\rho,\rho}(x) \equiv {3\over x^3}\int_0^x \xi_{\rho,\rho}(y) y^2 dy$.
The reason for using $\bar\xi_{\rho,\rho}$ instead of 
$\xi_{\rho,\rho}$ is that 
$\bar\xi_{\rho,\rho}$ is a better
indicator of nonlinearity than $\xi_{\rho,\rho}$.
The form of $D$ should be constrained at the linear end: $D(0)=1$.
As we will show below,
it turns out that
equation 28 fits results very well; i.e.,
fitting to numerical results indicates that $D=1$ is a good approximation
for the interested range in $\xi_{pk,pk}$. 
To be clear, we use equation 28 for all the subsequent
calculations of cluster correlation functions.

\begin{figure*}
\centering
\begin{picture}(400,300)
\psfig{figure=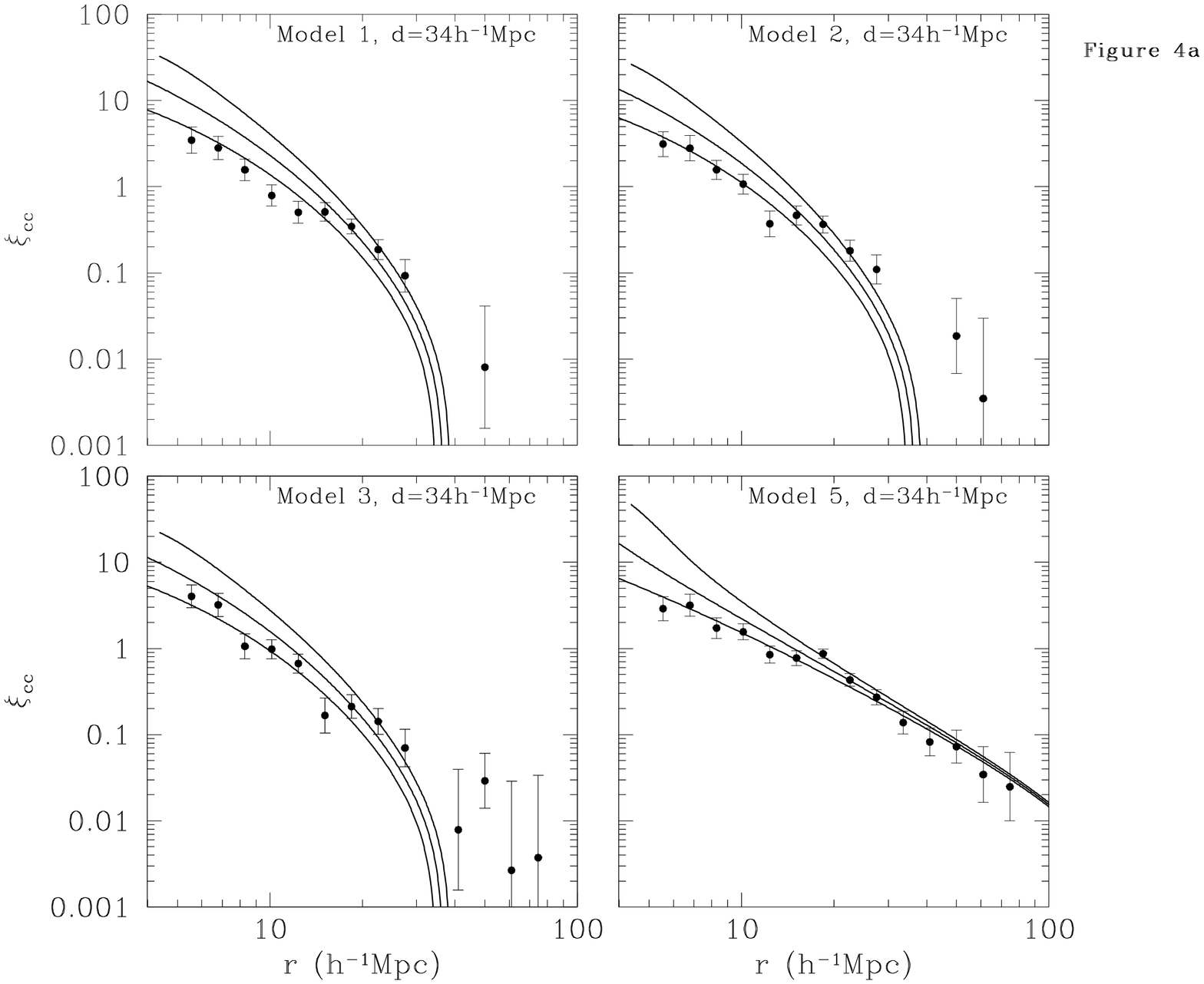,height=10.0cm,width=15.0cm,angle=-90.0}
\end{picture}
\centerline{(4a)}\vspace{0.1in}
\end{figure*}

\begin{figure*}
\centering
\begin{picture}(400,300)
\psfig{figure=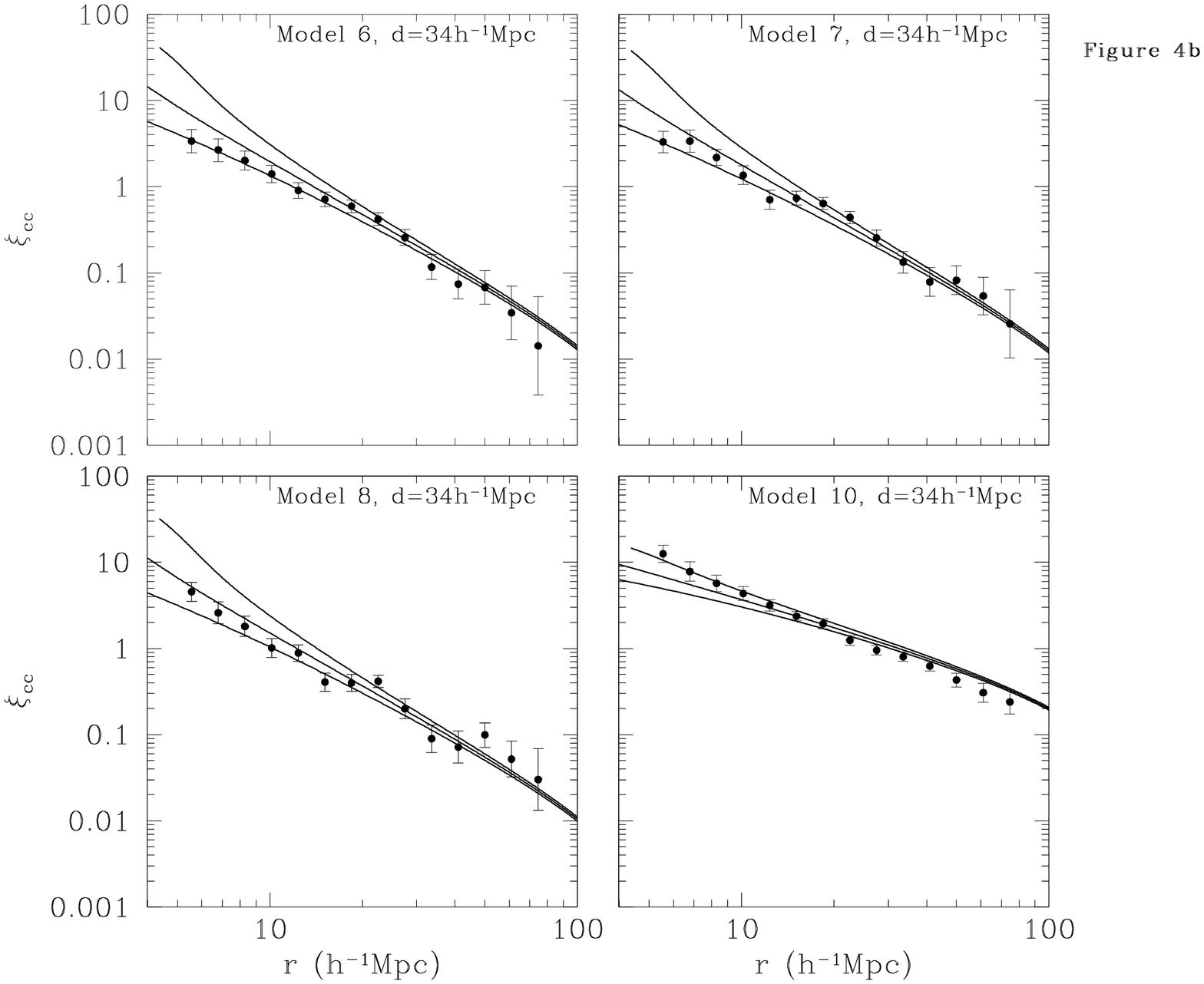,height=10.0cm,width=15.0cm,angle=-90.0}
\end{picture}
\centerline{(4b)}\vspace{0.1in}
\end{figure*}

\begin{figure*}
\centering
\begin{picture}(400,300)
\psfig{figure=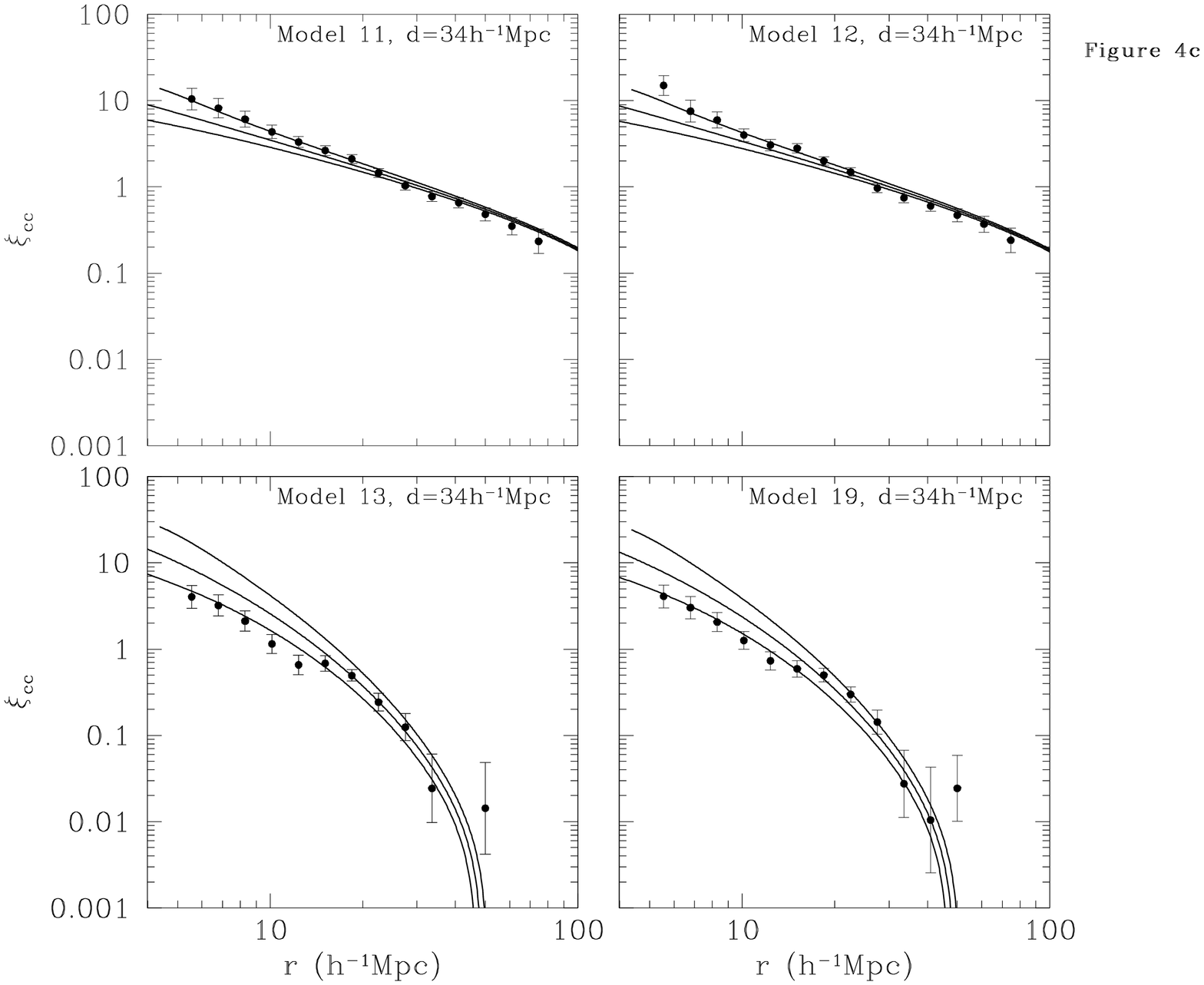,height=10.0cm,width=15.0cm,angle=-90.0}
\end{picture}
\centerline{(4c)}\vspace{0.1in}
\end{figure*}

\begin{figure*}
\centering
\begin{picture}(400,300)
\psfig{figure=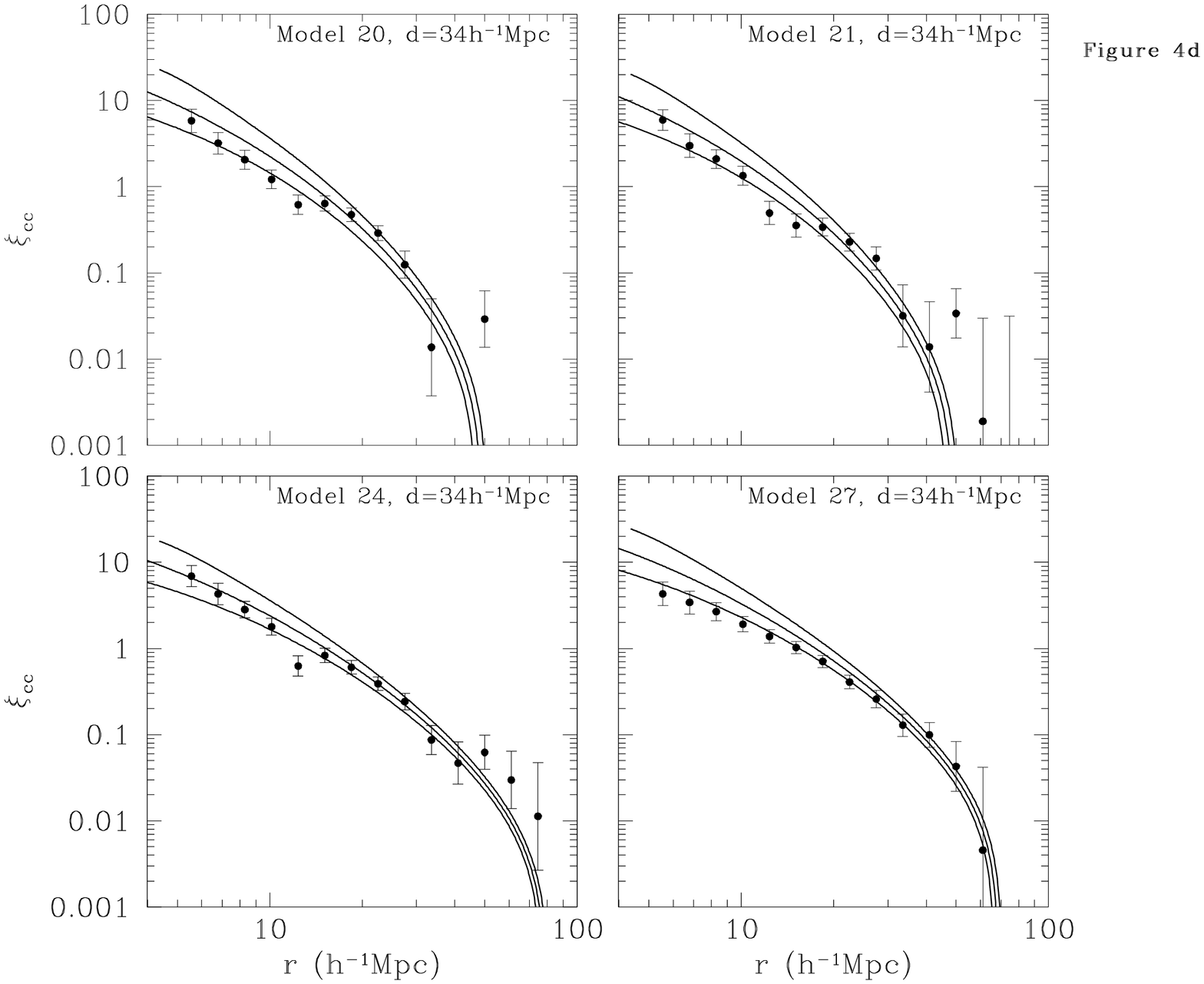,height=10.0cm,width=15.0cm,angle=-90.0}
\end{picture}
\centerline{(4d)}\vspace{0.1in}
\end{figure*}

\begin{figure*}
\centering
\begin{picture}(400,300)
\psfig{figure=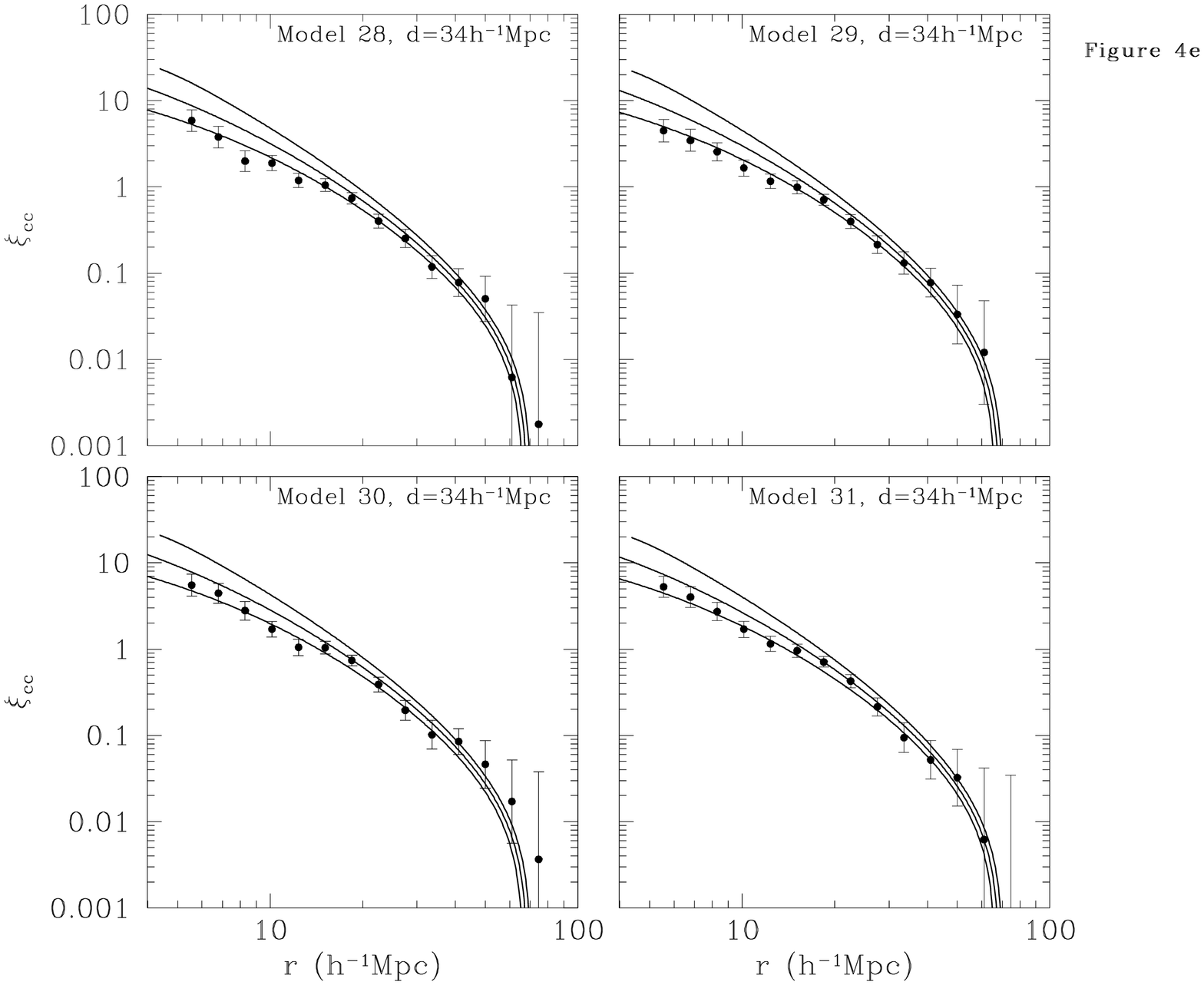,height=10.0cm,width=15.0cm,angle=-90.0}
\end{picture}
\centerline{(4e)}\vspace{0.1in}
\caption{
The cluster correlation functions
for clusters with mean separation of $34h^{-1}$Mpc, for 20 models
as indicated in the panels.
The errorbars are $1\sigma$ statistical.
}
\end{figure*}

We now compare the two-point peak-peak correlation
function, calculated using GPM described above,
to the cluster-cluster two-point correlation
function obtained from N-body simulations.
We compute the two-point correlation function from N-body
simulations using the following estimator:
\begin{equation}
\xi_{cc}(r)= {N_{CR}(r)\over N_{RR}(r)} - 1,
\end{equation}
\noindent 
where $N_{CR}(r)$ and $N_{RR}(r)$ are the 
number of pairs between clusters and random spatial points
and the number of pairs among 
random spatial points,
respectively, at separation $r\rightarrow r+\Delta r$.
The number of random spatial points for each realization
within the simulation box ($400h^{-1}$Mpc)
is chosen to be the same as the number of clusters in question.
A total of 100 random realizations are made.
The final $\xi_{cc}$ is averaged over the 100 estimations
and errorbars are estimated by 
separately computing the correlations for
each of the eight octants of each simulation box.

Figures (3a,b,c,d,e,f,g) show the results of 28 models
for clusters with mean separation of $55h^{-1}$Mpc.
Three of the remaining four models do not have
enough clusters whose masses exceed
$1.4\times 10^{14}h^{-1}\msun$
(the low mass cutoff for clusters found in the simulations).
Model 32 is not shown only to save space,
although
the goodness of the fit of GPM result
to the N-body result in Model 32 is comparable to that
of Model 15 (top left panel of Figure 3d).
Figures (4a,b,c,d,e) show the results of 20 models
for clusters with mean separation of $34h^{-1}$Mpc.
The models which are not shown, again,
do not have enough clusters with masses 
greater than $1.4\times 10^{14}h^{-1}\msun$,
except for Model 32.
The errorbars are $1\sigma$ statistical.
Three curves are shown in each panel for each model;
the middle curve is what is obtained using equation 28
and the top and bottom curves are obtained by 
adding $\pm 2h^{-1}$Mpc to each point 
of the middle curve in the x-axis.
We see that GPM works well
in the range of scales where $\xi_{cc}=0.1-2.0$
for all the models.
Some models fit for even larger ranges. 
In particular, the correlation length seems quite accurately
computable by GPM;
$r_0(GPM)\pm 2h^{-1}$Mpc 
[where $r_0(GPM)$ is the length
scale where the correlation computed by GPM (equation 28) is unity]
appears in agreement with the correlation length
computed from direct N-body simulations
for all the models studied here.

Several authors 
have used peaks in Gaussian density fields
to compute the cluster-cluster two-point correlation function.
Mann, Heavens, \&  Peacock (1993)
use a method developed by Bond \& Couchman (1988)
which combines the Gaussian peak formalism
with Zel'dovich approximation.
Their method is analytically tractable and fast.
Unfortunately, the accuracy of their method has not been 
carefully checked by N-body simulations and the two parameters
$r_f$ and $\delta_c$ (or $r_f$ and $\nu_t$)
are not fully deterministic.
It will be useful to apply the treatment of
$r_f$, $\alpha$ and $\delta_c$ here to their method.
Holtzman \& Primack (1993)
use a similar but
somewhat different formula (Bardeen, Bond, \& Efstathiou 1987)
than what is used here to compute the
cluster correlation function in some variants of CDM models.
Aside from the slightly different formula used to 
compute the statistical and dynamical correlation terms,
the primary difference between ours and theirs
lies in the treatment of $r_f$ and $\delta_c$.
In their work $r_f$ is determined by the mass of a cluster
in a rough way without taking into account the fact
that the virial radius of a collapsed object is
different from the radius within which observed mass 
is defined, which consequently also affects $\delta_c$.
It might be beneficial to check their analytic 
method against N-body simulations for a wide range of models,
before firm conclusions from detailed comparisons 
between models and observations can be drawn.
Croft \& Efstathiou (1994)
have checked the same formula as used here to 
compute the correlation function of clusters
using N-body simulations.
The difference between our method and theirs is that
$r_f$ and $\delta_c$ in their approach are
only roughly guessed based on a consideration of 
the approximate mass of clusters which determines $r_f$ and 
the spherical collapse model which determines $\delta_c=1.69$.
Their Gaussian peak method agrees well with their N-body 
results for the poorer clusters ($d_c=15h^{-1}$Mpc) 
in all three models ($\Omega_0=1$ model, $\Omega_0=0.2$ and
$\Lambda_0=0.8$ model, and $\Omega_0=0.2$ and $\Lambda_0$ model)
but the correlation functions from their
Gaussian method are
significantly lower than those from N-body simulations
for richer clusters ($d_c=35,50h^{-1}$Mpc).
The behavior of their results can be explained 
by the fact that they choose $r_f$ 
to be independent of $M_A$ (or $d_c$, the mean separation
of clusters under consideration; 
a larger $d_c$ corresponds to richer, more massive clusters),
whereas in our case $r_f$ correlates with $M_A$.
Consequently, the correlations of richer clusters from the
Gaussian method of Croft \& Efstathiou (1994)
are underestimated.
The primary drawback of all these previous studies
is the inability to uniquely
identify a set of peaks with a set of clusters of interest.
Again, this traces to the ambiguity of choosing $r_f$ and $\nu_t$.
Our method eliminates this ambiguity entirely and the 
results seem satisfactory.

\subsection{Calibrating Press-Schechter Formalism}

We have shown in \S 2.1 and \S 2.2
that GPM works quite well for computing both
the cluster mass and correlation functions.
It is tempting to try the same arguments 
on the widely used Press-Schechter (1974; PS) formalism.
We recall that 
the essential ingredient in GPM 
is the introduction of two parameters:
$\alpha$ and $\delta_c$.
Fitting results from GPM to N-body results
fixes $\alpha$ to be a constant of $2.3$
and $\delta_c$ as a function of $\Omega_z$
(equation 27).
The fitted value of $\alpha$ is consistent
with both the theoretical
work by Navarro, Frenk \& White (1996)
and observations
(Carlberg \etal 1996; Fischer \etal 1997).
So, we will adopt the same $\alpha$ derived from
GPM to calibrate the PS formalism.
But we will adjust $\delta_c$ 
by comparing PS results to N-body results.

\begin{figure*}
\centering
\begin{picture}(400,300)
\psfig{figure=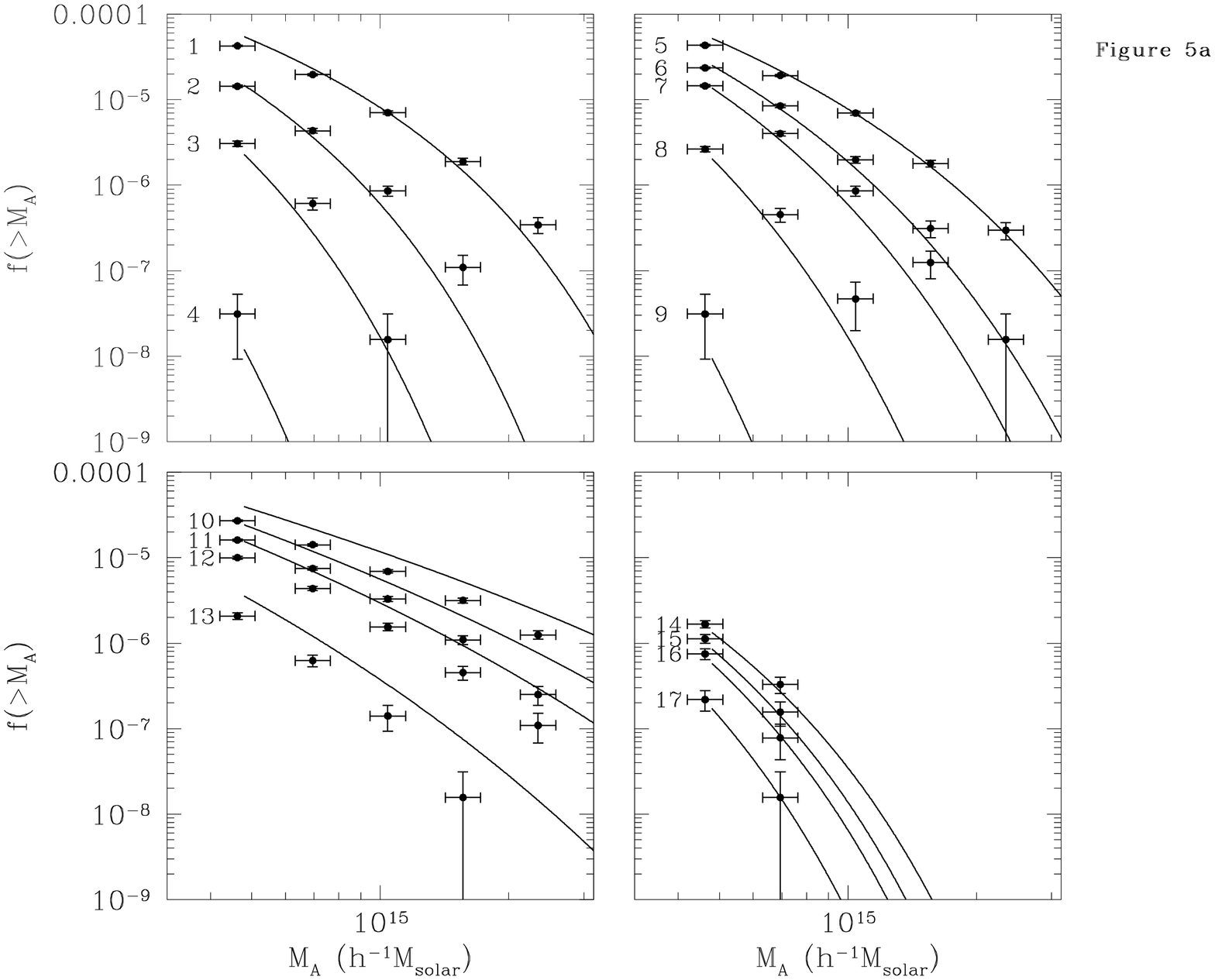,height=10.0cm,width=15.0cm,angle=-90.0}
\end{picture}
\centerline{(5a)}\vspace{0.1in}
\end{figure*}

\begin{figure*}
\centering
\begin{picture}(400,300)
\psfig{figure=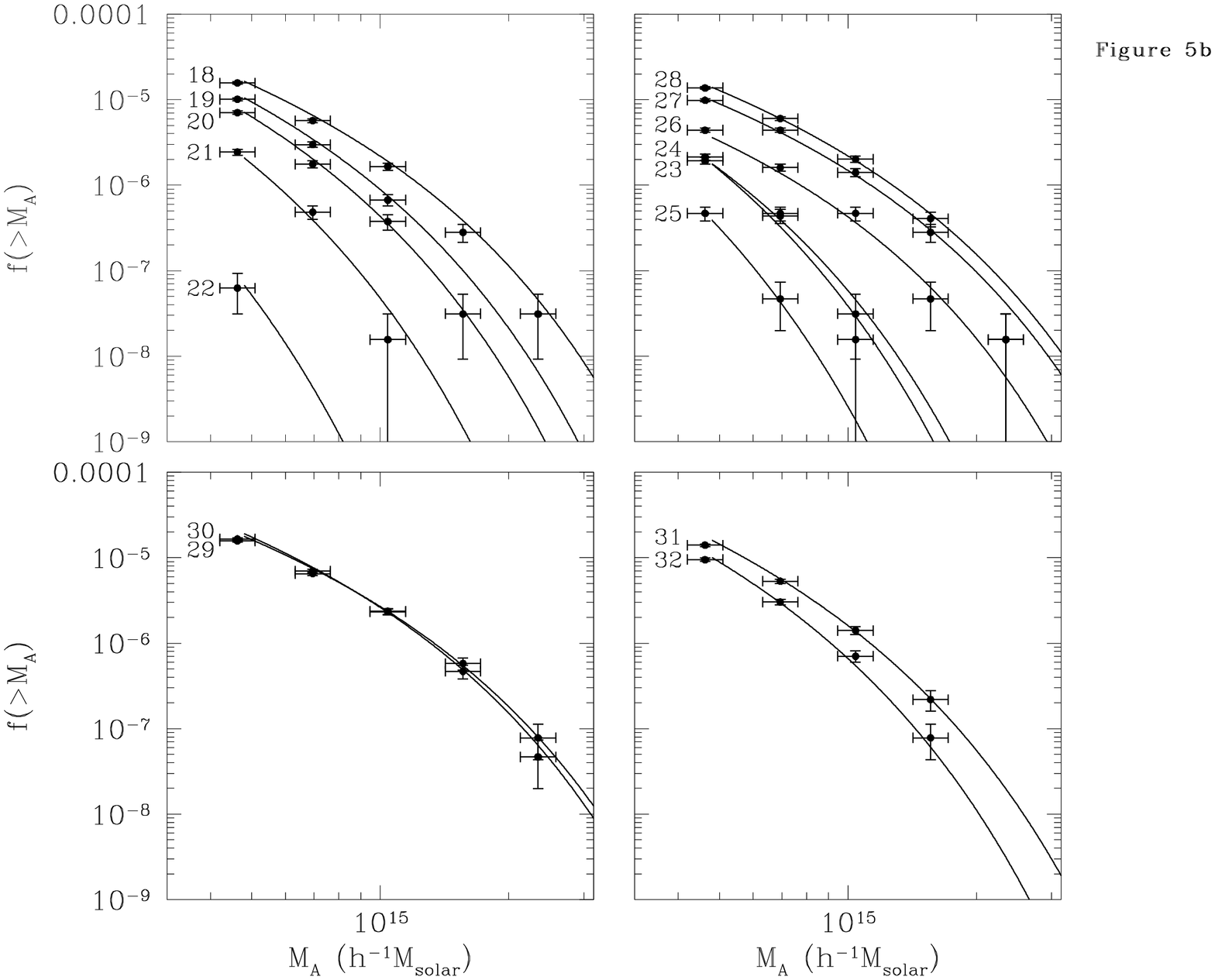,height=10.0cm,width=15.0cm,angle=-90.0}
\end{picture}
\centerline{(5b)}\vspace{0.1in}
\caption{
Mass functions of various models.
The simulation results are shown
as symbols with horizontal errorbars
being the uncertainties in the mass determination (15\%)
and the vertical errorbars being the statistical $1\sigma$ errorbars
for the number of clusters.
The solid curves are the results from Press-Schechter method.
}
\end{figure*}

The basic PS ansatz results in
the differential {\it virialized halo mass} function as:
\begin{eqnarray}
n(M_v)dM_v=\hskip -0.6cm &&-\sqrt{2\over \pi}{\bar\rho\over M_v}{\delta_c\over\sigma_0^2(M_v)} {d\sigma_0(M_v)\over dM_v}\nonumber \\
\hskip -0.6cm &&\exp(-{\delta_c^2\over 2\sigma_0^2(M_v)}) dM_v,
\end{eqnarray}
\noindent 
where $\bar\rho$ is the mean density of the universe
at the redshift under consideration.
Substituting $M_v$ by $M_A$ using equation 14 with $\alpha=2.3$
gives
\begin{eqnarray}
n(M_A)dM_A=\hskip -0.6cm&&-4.945\times 10^{15} r_A^{0.913} \Omega_z^{0.173} \Omega_0^{1.304}\nonumber \\
\hskip -0.6cm&&M_A^{-1.304}{\delta_c\over\sigma_0^2(M_A)} {d\sigma_0(M_A)\over dM_A}\nonumber\\
\hskip -0.6cm&&\exp(-{\delta_c^2\over 2\sigma_0^2(M_A)}) dM_A.
\end{eqnarray}
\noindent 
To calculate $\sigma_0(M_A)$ in the above equation
we use the {\it Gaussian} smoothing window  with
the radius determined by equation 13.
The original PS formalism was based on
the sharp k-space filter,
but it has been shown subsequently by many authors
that Gaussian filter works at least as well.
The additional virtue of a Gaussian window is
that it guarantees a convergent integral for $\sigma_0$
for any plausible power spectrum.
Now, the only parameter left undetermined is $\delta_c$,
which will be fixed by comparing to N-body results.
We find that the best overall fit of PS results 
to N-body results is obtained,
if
\begin{equation}
{\delta_c=\cases{&\hskip -0.5cm$1.23-0.05(1.0-\Omega_z)\quad\quad\hbox{for}\quad \Lambda_z=0$ \cr 
&\hskip -0.5cm$1.23-0.01(1.0-\Omega_z)\quad\quad \hbox{for}\quad \Omega_z+\Lambda_z=1$.\cr}}
\end{equation}
\noindent 
The results are shown in Figure 5.
We see that PS fits N-body results quite well
except for the $P_k=k^{-2}$ models (models 10,11,12,13).
The PS results for all the $\Omega_0=1$
models except
the $P_k=k^{-2}$ models
appear to be somewhat above the N-body results
at the low mass end ($\sim 5\times 10^{14}h^{-1}\msun$)
and somewhat below 
the N-body results
at the high mass end ($\sim 2.5\times 10^{15}h^{-1}\msun$).
On the other hand,
the PS results for all the $P_k=k^{-2}$ models
are significantly above the N-body results.
So, there is no room for further adjustments of $\delta_c$
to achieve better overall fits,
at least for Gaussian smoothing windows.

Note that $\delta_c$ ($\sim 1.23$) is smaller than $1.67$,
which is in the expected direction
because a smoother, Gaussian smoothing window is used here.
$1.23$ is also somewhat smaller than 
that given by Klypin \etal (1995), who give
$\delta_c=1.40$ for a Gaussian smoothing window in
the context of damped Lyman alpha systems.
But Klypin \etal also argue
that $\delta_c$ could be as low as $1.3$,
were waves longer and shorter than those present in the 
simulation box included.
We suspect that $\delta_c$ also depends on the 
shape of the power spectrum in a way that is
analogous to the difference between different smoothing windows:
a steeper power spectrum
(i.e., $n$ being smaller with
$P_k=k^{n}$), which conspires to 
form a sharp k-space filter like that
used in the original derivation of PS,
requires a larger $\delta_c$, while a flatter power spectrum
requires a smaller $\delta_c$.
With this conjecture, the trend that 
applications of PS to smaller cosmic objects
tend to require larger $\delta_c$ would have been predicted,
since CDM-like spectra have a slowly bending shape which
is flatter at small $k$ (i.e., for larger systems)
and steeper at large $k$ (i.e., for smaller systems).
This conjecture seems to be borne out
in the subsequent analyses, as best summarized as in
equation 36, where the dependence of $\sigma_8$
on $\Gamma$ is consistent with the above hypothesis.
This issue will be addressed elsewhere in more detail.

While PS works well for CDM-like models
for computing halo mass functions, consistent
with earlier works (Efstathiou \& Rees 1988;
WEF; ECF),
it seems that GPM fits somewhat better the N-body results
for CDM-like models and also works for other models tested.
An additional advantage is that GPM allows for 
a determination of the correlation function as well.
Therefore, in subsequent calculations we will use GPM,
if deemed appropriately applicable.

\section{Various Factors That Affect Cluster Mass Function}

\begin{figure*}
\centering
\begin{picture}(400,300)
\psfig{figure=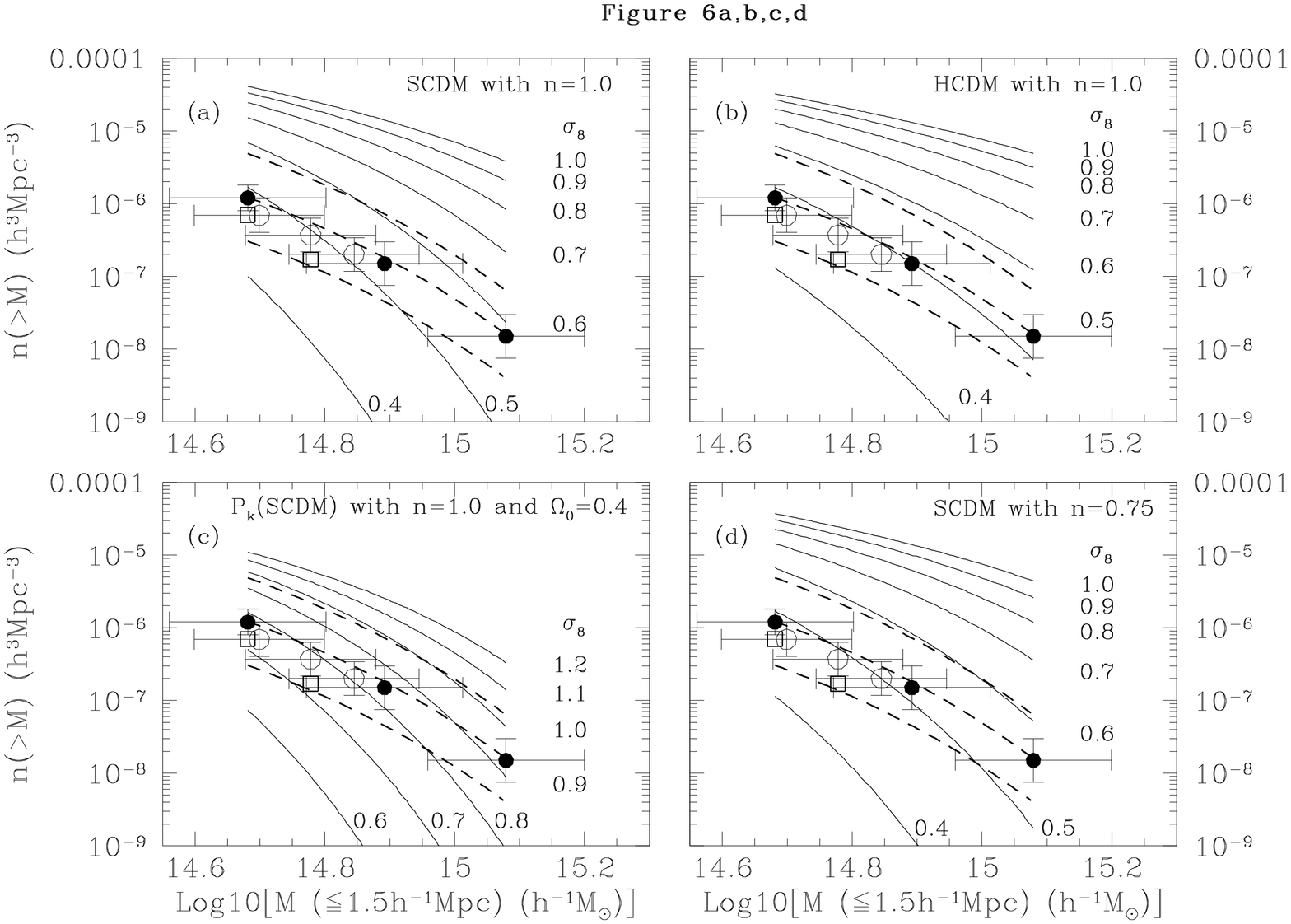,height=10.0cm,width=15.0cm,angle=-90.0}
\end{picture}
\centerline{(6a,b,c,d)}\vspace{0.1in}
\end{figure*}

\begin{figure*}
\centering
\begin{picture}(400,300)
\psfig{figure=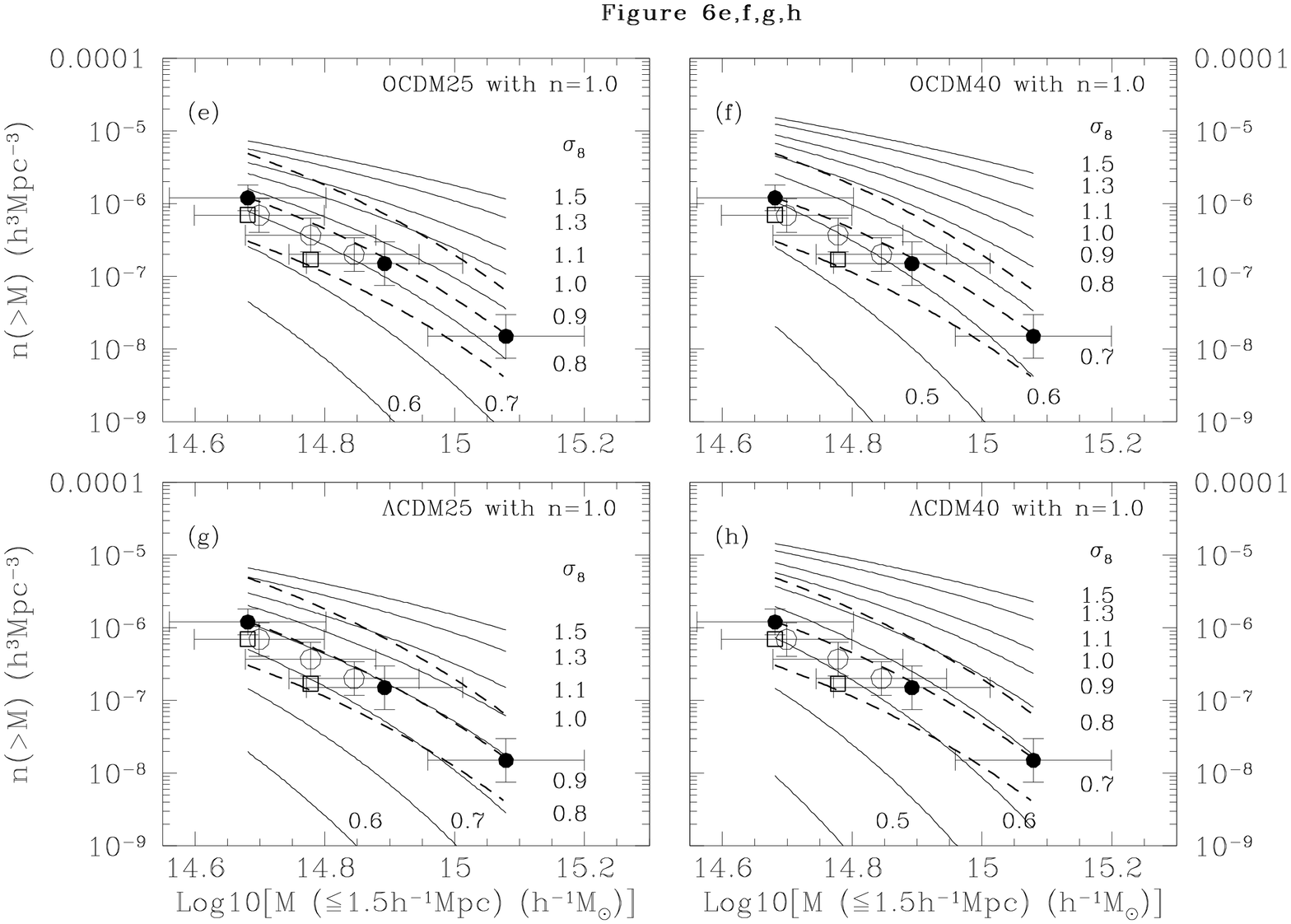,height=10.0cm,width=15.0cm,angle=-90.0}
\end{picture}
\centerline{(6e,f,g,h)}\vspace{0.1in}
\caption{
Cluster mass functions for six models at several
different normalization amplitudes of the power spectra.
Also shown as symbols are the observations adopted
from BC,
and as three dashed curves are the fits to the symbols.
The solid dots represent the cluster mass function 
from Abell cluster catalog,
the open squares from the 
Edinburgh-Durham Cluster Catalog (Lumsden \etal 1992),
and the open circles  
from the observed temperature function 
of Henry \& Arnaud (1991). 
The middle dashed curve is computed by equation 35, $n_{fit}(>M)$,
which represents the mean value of
the observed mass function well.
The top and bottom dashed curves
are $4n_{fit}(>M)$ and $0.25n_{fit}(>M)$, 
respectively.
}
\end{figure*}

It is worthwhile to understand what factors are relevant
for the cluster mass function.
We begin by showing 
the cluster mass functions for six variants of the standard CDM model
(Table 2 below),
as indicated in panels (a,b,e,f,g,h) of Figure 6,
at different normalization amplitudes ($\sigma_8$).
We will return to Table 2 in \S 4 to discuss 
the various models in detail.
The primordial power spectrum index
is assumed to be $n=1$ for the shown models in panels (a,b,e,f,g,h).
Also shown as symbols are the observations adopted
from BC,
and as three dashed curves are the fits to the symbols.
The middle dashed curve is
\begin{eqnarray}
n_{fit}(>M_A)=\hskip -0.6cm&&2.7\times 10^{-5}(M_A/2.1\times 10^{14})^{-1}\nonumber\\
\hskip -0.6cm&&\exp(-M_A/2.1\times 10^{14}),
\end{eqnarray}
\noindent 
where $n_{fit}$
is in $h^3$Mpc$^{-3}$ and
$M_A$ is the cluster mass within the Abell radius
in $h^{-1}\msun$.
This curve seems to represent the mean value of
the observed mass function well (note that equation 35
is slightly different from the fitting formula in BC).
The top and bottom dashed curves
are $4n_{fit}(>M_A)$ and $0.25n_{fit}(>M_A)$.
It is difficult to estimate the errorbars of the observed
mass function.
The top and bottom dashed curves 
are intended to serve
as $2\sigma$ upper and lower bounds (in the vertical
axis) of the observed mass function within the indicated mass range,
which we deem to be conservative.
Subsequent presentations and explanations will follow this assertion.

In all cases, we see that the cluster mass function 
becomes progressively steeper at the high mass end
as the amplitude of the density perturbations decreases.
The physics behind this is simple to understand.
As the amplitude of fluctuations decreases,
the required height of the density peaks for 
clusters with a given mass increases.
Since the abundance of the high peaks at the very high end
drops exponentially,
the mass function steepens as
the amplitude of fluctuations decreases.
Note that there is only a narrow range in $\sigma_8$
where the model mass function lies within the $2\sigma$
limits in the mass range from $4.8\times 10^{14}h^{-1}\msun$
to $1.2\times 10^{15}h^{-1}\msun$.

Although it is clear that the amplitude of the power spectrum
sensitively determines the abundance of the clusters,
as seen by comparing different curves within each panel,
it is not yet clear how big the effect of $\Omega_0$ 
on the mass function is.
In Figure 6, panel (c) is similar to panel (a) 
with only one change: $\Omega_0=0.4$ instead of $\Omega_0=1.0$.
Note that the power spectrum used in panel (c) is identical
to that used in panel (a).
We see $\Omega_0$ has a significant effect on the mass function.
For example, the $\sigma_8=1.0$ model in panel (c) has comparable
mass function to the $\sigma_8=0.6$ model in panel (a),
but about two orders of magnitudes lower than
the $\sigma_8=1.0$ model in panel (a).

Next, we examine the effect of the shape of the power
spectrum on the shape of the mass function.
In Figure 6 panel (d) is similar to panel (a)
but with $n=0.75$ instead of $n=1.0$.
Comparing panel (d) with
panel (a) illustrates the sensitivity of the cluster mass function
on the shape of power spectrum.
The model shown in panel (d) has substantially more power
on large scales ($100-300h^{-1}$Mpc)
than the model shown in panel (a).
This difference results in flatter
mass functions in panel (d) than in panel (a), 
especially for the cases with lower $\sigma_8$.
A point that we would like to make here is 
that the cluster mass function depends not only 
on $\Omega$, $\Lambda$ and $\sigma_8$, 
but on $P_k$ in a nontrivial, albeit relatively weak, way,
especially at the high mass end of the mass function.
For example, at $1.2\times 10^{15}h^{-1}\msun$, 
the model shown in panel (d) has about a factor of
four more clusters than the model shown in panel (a),
with the only difference between the two models
being a slight tilt of the power spectrum 
[$n=0.75$ in (d) vs $n=1.0$ in (a)].

Panels (e,g) and panels (f,h)
show two pairs of low density models, one being open and the
other being flat with a cosmological constant but
with a same $\Omega_0$ in each pair.
One thing to note is that,
other things being fixed, 
the cosmological constant does not
make much difference in terms of the shape of the mass function.
However, for a fixed $\sigma_8$, 
the mass function in the model with a cosmological
constant is systematically lower than that
of the model without a cosmological
constant;
the difference becomes larger at lower amplitudes.

Summarizing the above results we see
that the factors that effect the cluster mass function
in order of decreasing importance 
are $\sigma_8$, $\Omega_0$, $P_k$ and $\Lambda_0$.
The ordering of the last two factors is somewhat more complex;
the $P_k$ factor is more significant at the high mass end,
whereas the $\Lambda_0$ factor affects rather 
uniformly across-the-board.
It is therefore clear that the tightest
constraint may be obtained for $\sigma_8$
for a given model with $\Omega_0$, $\Lambda_0$ and $P_k$ 
being specified.

\section{Normalizing All CDM Models}

\begin{deluxetable}{cccccccccc} 
\tablewidth{0pt}
\tablenum{2}
\tablecolumns{8}
\tablecaption{Six variants of CDM models} 
\tablehead{
\colhead{Model Family} &
\colhead{$\Omega_c$} &
\colhead{$\Omega_h$} &
\colhead{$\Lambda_0$} &
\colhead{H$_0$} &
\colhead{$\Omega_b$} &
\colhead{$t_{age}~$(Gyrs)}}

\startdata
A & $0.936$ & $0.000$ & $0.000$ & $55$ & $0.064$ & $11.8$ \nl 
B & $0.736$ & $0.200$ & $0.000$ & $55$ & $0.064$ & $11.8$ \nl 
C & $0.220$ & $0.000$ & $0.000$ & $65$ & $0.030$ & $12.4$ \nl 
D & $0.346$ & $0.000$ & $0.000$ & $60$ & $0.054$ & $12.7$ \nl 
E & $0.220$ & $0.000$ & $0.750$ & $65$ & $0.030$ & $15.2$ \nl 
F & $0.346$ & $0.000$ & $0.600$ & $60$ & $0.054$ & $14.5$ \nl 
\enddata
\end{deluxetable}

In \S 2.1 and 2.2
we have shown that the Gaussian peaks of cluster size
indeed form clusters of galaxies and appropriately identified
peaks reproduce both cluster mass function and
cluster two-point correlation function accurately.
We now use the observed
zero redshift rich cluster abundance and the COBE observation
to constrain six representative variants of the standard CDM
model, which are of current interest.
The models are listed in Table 2.

The baryon densities for Models C,E
are computed using $\Omega_b h^2=0.0125$ (Walker \etal 1991),
and for Models A,B,D,F 
using $\Omega_b h^2=0.0193$ (Burles \& Tytler 1997).
These choices of $\Omega_b$ for the models
serve to maximize the viability of each model
with respect to the observed gas fraction
in X-ray clusters of galaxies (White \etal 1993;
Lubin \etal 1996;
Danos \& Pen 1998, $\rho_{gas}/\rho_{tot}=(0.053\pm 0.004)h^{-3/2}$).
The power spectrum transfer functions for all the models
are computed using the CMBFAST code developed by Seljak and Zaldarriaga.
The choice of the Hubble constant is made for each model such that
each model is consistent with current measurements of
the Hubble constant.
It appears to be a consensus 
that $H_0(obs)=65\pm 10$km/s/Mpc can account for
the distribution of the current data from various measurements
(see, e.g., Trimble 1997),
except for those from Sunyaev-Zel'dovich observations
(for a discussion of 
a reconciliation of this difference, see Cen 1998).
Another consideration is that
the age constraint
from latest globular cluster observations/interpretations
(c.f., Salaris, Degl'Innocenti, \& Weiss 1997)
is not violated.

The SCDM model (\# 1 in Table 2)
is the standard CDM with critical
density.
The mixed hot and cold dark matter model (HCDM)
has
critical density but with $20\%$ of the mass in
light massive neutrinos (two species of equal mass neutrinos
are assumed).
The next two models, OCDM25 and OCDM40 (\# 3 and 4),
are open models
with matter density of $\Omega_0=0.25$ and $0.40$, respectively.
The last two models, $\Lambda$CDM25 and $\Lambda$CDM40
(\# 5 and 6), are spatially flat models with the addition of 
a cosmological constant but with otherwise similar parameters
to the two open models.
Note that for all models, $\sigma_8$ and $n$ are yet to be specified,
as will be shown below, by
normalizing each model to both COBE and 
zero redshift cluster abundance.

GPM makes possible economically
sample large (four dimensional) parameter space spanned
by the uncertainties in $\Omega_0$, $\Lambda_0$, $\sigma_8$ and $P_k$
(the Hubble constant constitutes
yet another dimension of uncertainty but its effect can be absorbed 
into $P_k$ in this case since the final quantities
can all be expressed in units of $h$).
The dependence on $\Omega_b$ is secondary and ignored here.
If the cluster mass function of a model at any 
mass does not lie within the 
range delimited by the upper and lower dashed curves
in Figure 6
in the mass range from $4.8\times 10^{14}h^{-1}\msun$
to $1.2\times 10^{15}h^{-1}\msun$,
we conclude that the model is ruled out 
at $2\sigma$ confidence level
(this adoption of confidence level on the mass function
is somewhat crude but we think it is perhaps conservative).
Thus, we can constrain the 
($\Omega_0$, $\Lambda_0$, $\sigma_8$, $P_k$) four dimensional
parameter space by {\it  the cluster mass function}.
The other, tight constraint currently available 
is from the COBE observations of the cosmic microwave
background fluctuations on large scales.
We will combine these two observations to constrain the models.

\begin{figure*}
\centering
\begin{picture}(400,300)
\psfig{figure=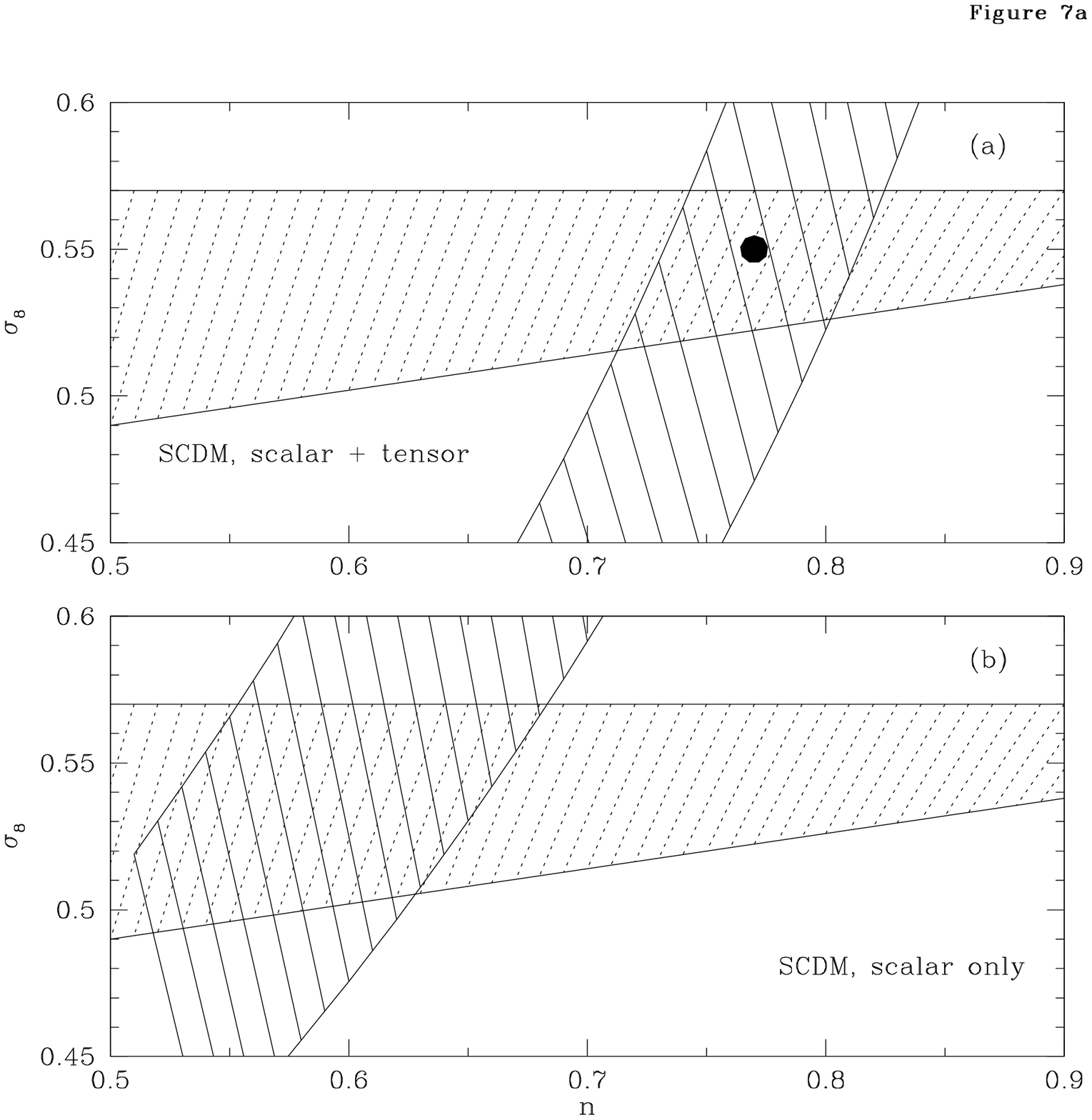,height=10.0cm,width=15.0cm,angle=0.0}
\end{picture}
\centerline{(7a)}\vspace{0.1in}
\end{figure*}

\begin{figure*}
\centering
\begin{picture}(400,300)
\psfig{figure=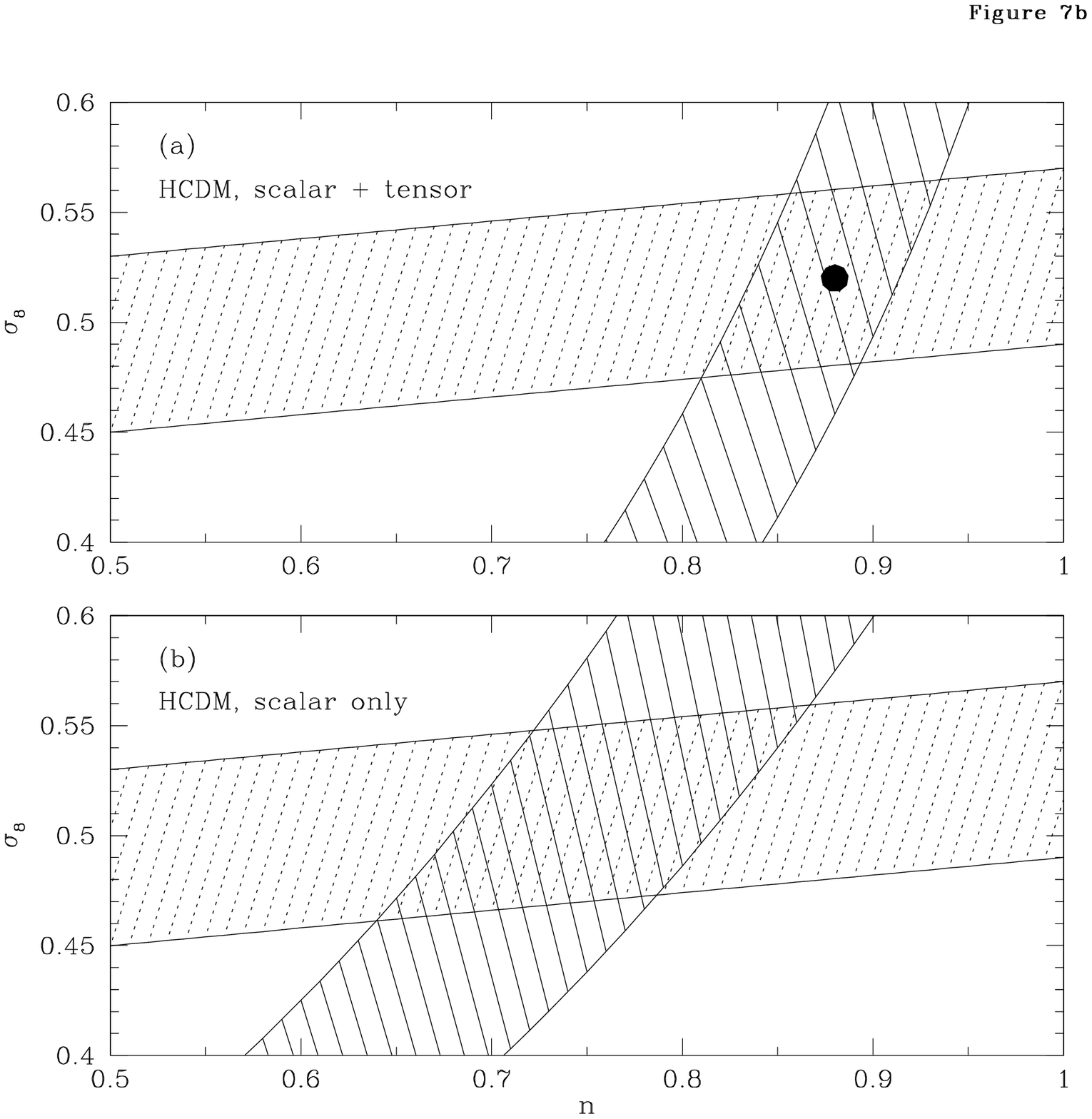,height=10.0cm,width=15.0cm,angle=0.0}
\end{picture}
\centerline{(7b)}\vspace{0.1in}
\end{figure*}

\begin{figure*}
\centering
\begin{picture}(400,300)
\psfig{figure=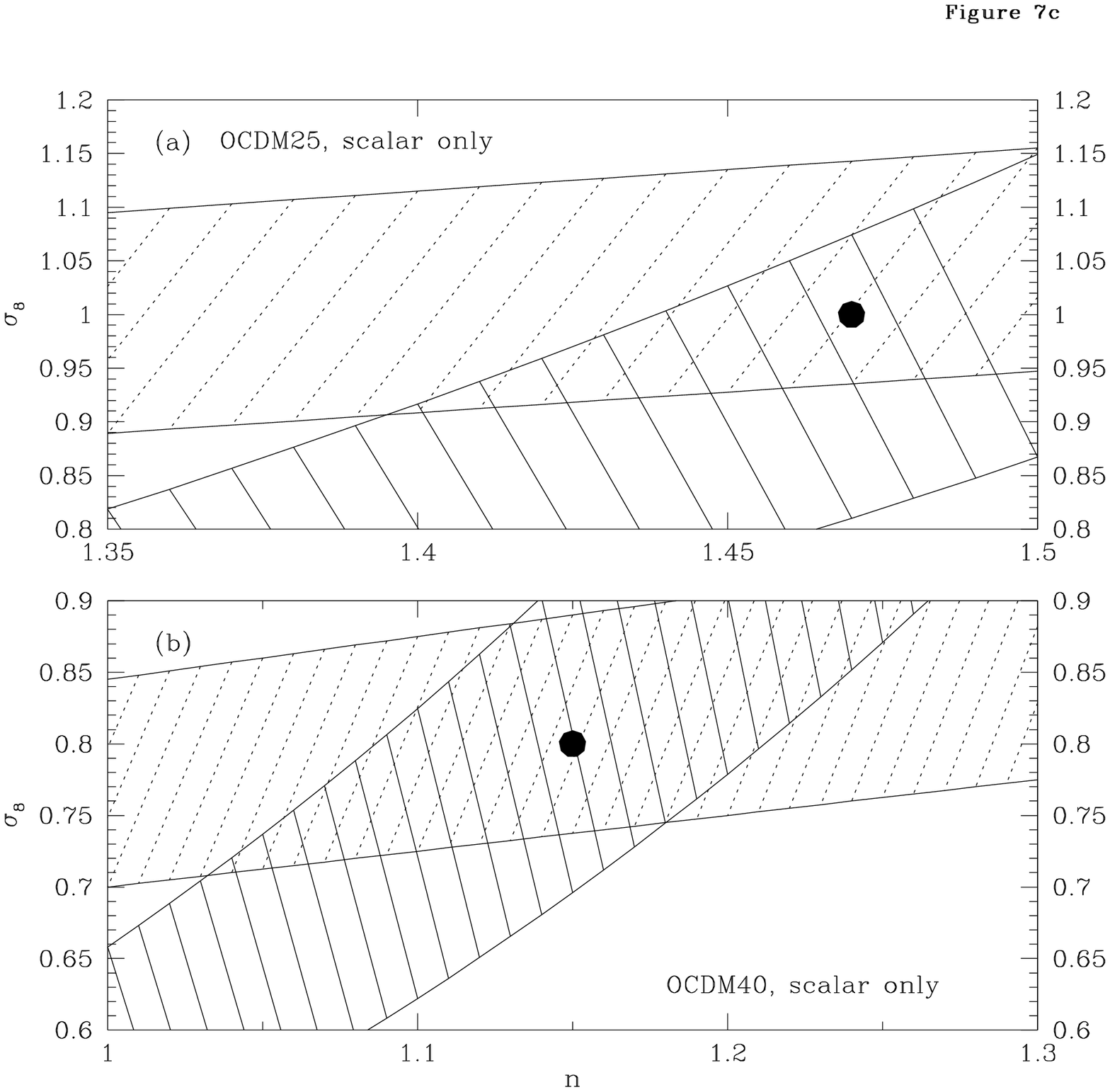,height=10.0cm,width=15.0cm,angle=0.0}
\end{picture}
\centerline{(7c)}\vspace{0.1in}
\end{figure*}

\begin{figure*}
\centering
\begin{picture}(400,300)
\psfig{figure=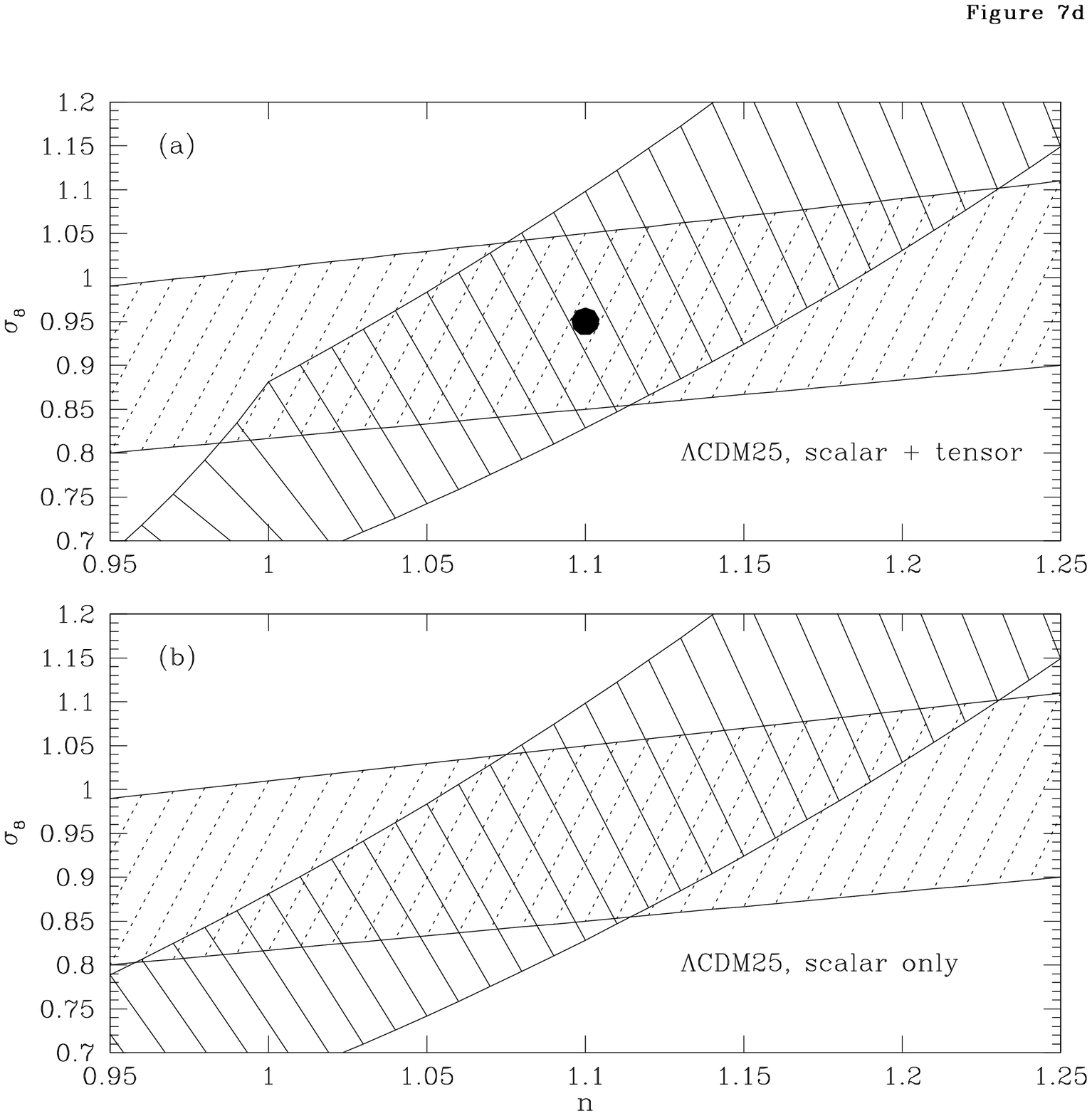,height=10.0cm,width=15.0cm,angle=0.0}
\end{picture}
\centerline{(7d)}\vspace{0.1in}
\end{figure*}

\begin{figure*}
\centering
\begin{picture}(400,300)
\psfig{figure=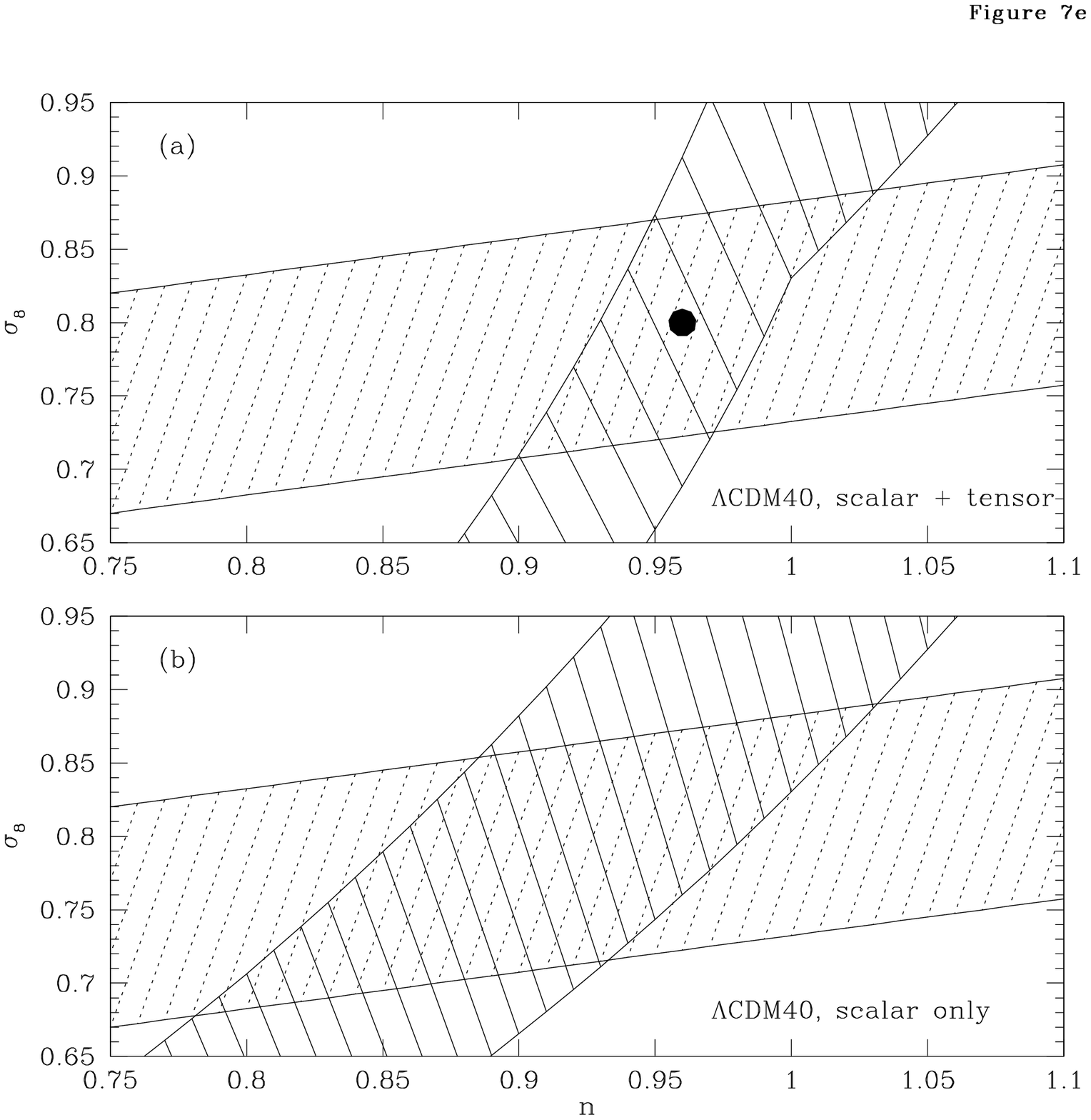,height=10.0cm,width=15.0cm,angle=0.0}
\end{picture}
\centerline{(7e)}\vspace{0.1in}
\caption{
The constrained two-dimensional parameter space ($\sigma_8,n$)
for the six models as tabulated in Table 2.
The dotted hatched regions are the permitted
space in this two-dimensional parameter plane
whose cluster mass functions
agree with observations at the $2\sigma$ confidence level.
The solid hatched regions
are the constraints provided
by COBE on large scales,
where the spread in the vertical axis ($\sigma_8$, $\pm 14\%$, $2\sigma$)
is due to the statistical uncertainty of the COBE
measurements (Bunn \& White 1997).
For the four spatially flat models (models 1,2,5,6)
two cases, with and without
the tensor mode (gravitational wave) contribution
to the CMB fluctuations on COBE scales
are considered (see text).
For the remaining two models (open models; models 3,4 of Table 2)
we only consider the case
without the gravitational wave contribution.
}
\end{figure*}

\begin{deluxetable}{cccccccccc} 
\tablewidth{0pt}
\tablenum{3}
\tablecolumns{6}
\tablecaption{Constraints on $n$, $\sigma_8$ and EP for the six variants of Table 2} 
\tablehead{
\colhead{Model Family} &
\colhead{Tensor mode} &
\colhead{$n$ range} &
\colhead{$\sigma_8$ range} &
\colhead{EP range}}

\startdata
A & yes & $0.72-0.82$ & $0.52-0.57$ & $1.01-1.05$ \nl 
A & no  & $0.45-0.68$ & $0.49-0.52$ & $1.07-1.20$ \nl 
B & yes & $0.79-0.92$ & $0.45-0.54$ & $1.44-1.53$ \nl 
B & no  & $0.61-0.84$ & $0.43-0.53$ & $1.49-1.65$ \nl 
C & no  & $1.39-1.49$ & $0.91-1.14$ & $1.11-1.16$ \nl 
D & no  & $0.98-1.23$ & $0.64-0.84$ & $1.25-1.40$ \nl 
E & yes  & $0.98-1.23$ & $0.81-1.10$ & $1.25-1.40$ \nl 
E & no   & $0.96-1.23$ & $0.80-1.10$ & $1.25-1.42$ \nl 
F & yes  & $0.87-0.99$ & $0.65-0.82$ & $1.40-1.47$ \nl 
F & no   & $0.74-0.98$ & $0.62-0.81$ & $1.40-1.56$ \nl 
COBE(obs) & - & $1.20\pm 0.60$ ($2\sigma$) & - & - \nl 
LSS(obs) & - & - & - & $1.30\pm 0.30$ ($2\sigma$) \nl 
\enddata
\end{deluxetable}

\begin{deluxetable}{cccccccccc} 
\tablewidth{0pt}
\tablenum{4}
\tablecolumns{10}
\tablecaption{Six COBE and Cluster-Normalized CDM Models} 
\tablehead{
\colhead{Model} &
\colhead{$H_0$} &
\colhead{$n$} &
\colhead{$\Omega_c$} &
\colhead{$\Omega_h$} &
\colhead{$\Lambda_0$} &
\colhead{$\Omega_b$} &
\colhead{$\sigma_8$}}

\startdata
tCDM & $55$ & $0.77$ & $0.936$ & $0.00$ & $0.0$ & $0.064$ & $0.55$ \nl 
HCDM & $55$ & $0.88$ & $0.736$ & $0.20$ & $0.0$ & $0.064$ & $0.52$ \nl 
OCDM25 & $65$ & $1.47$ & $0.220$ & $0.00$ & $0.0$ & $0.030$ & $1.00$ \nl 
OCDM40 & $60$ & $1.15$ & $0.346$ & $0.00$ & $0.0$ & $0.054$ & $0.80$ \nl 
$\Lambda$CDM25 & $65$ & $1.10$ & $0.220$ & $0.00$ & $0.75$ & $0.030$ & $0.95$ \nl 
$\Lambda$CDM40 & $60$ & $0.96$ & $0.346$ & $0.00$ & $0.60$ & $0.054$ & $0.80$ \nl 
\enddata
\end{deluxetable}

\begin{figure*}
\centering
\begin{picture}(400,300)
\psfig{figure=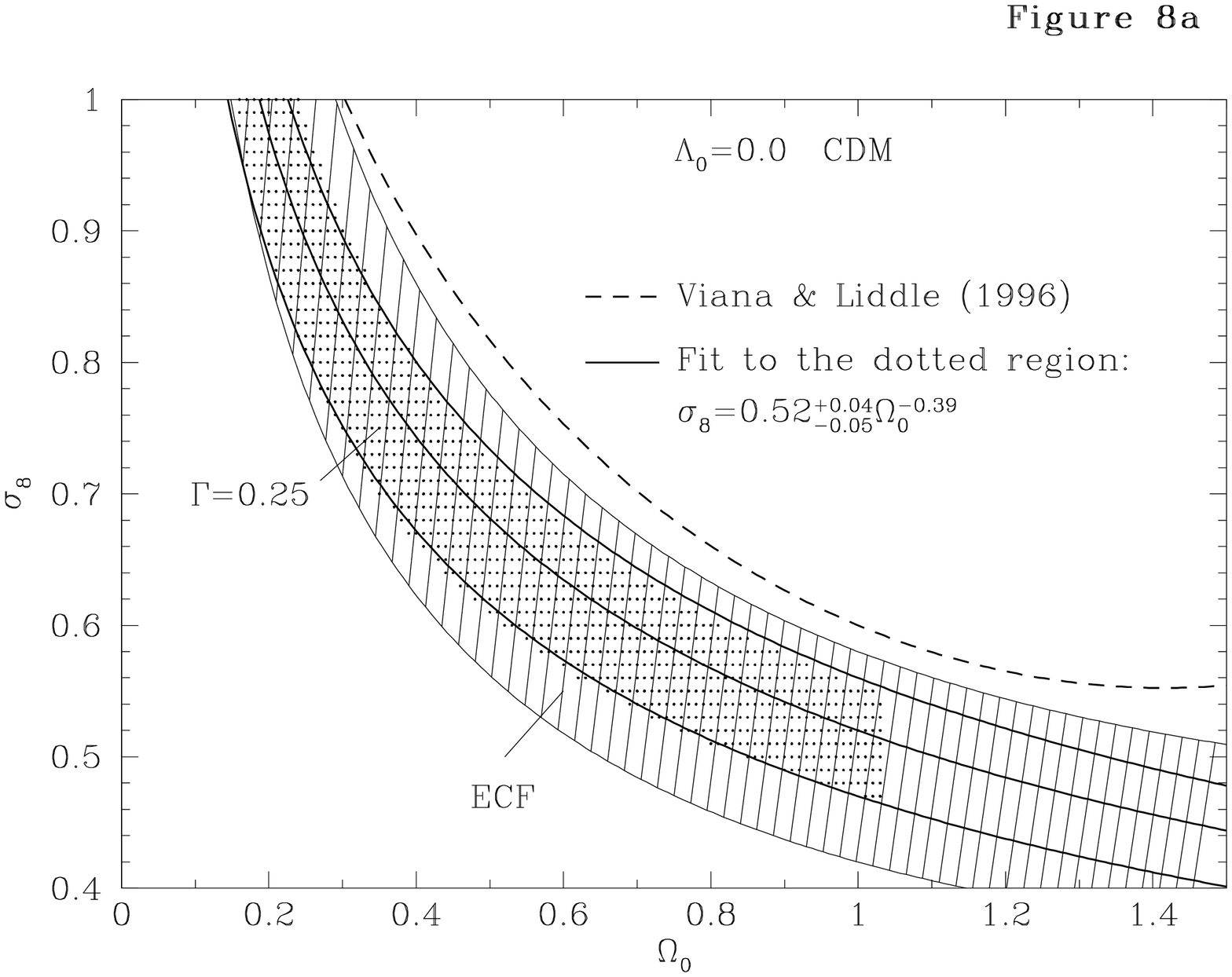,height=10.0cm,width=15.0cm,angle=-90.0}
\end{picture}
\centerline{(8a)}\vspace{0.1in}
\end{figure*}

\begin{figure*}
\centering
\begin{picture}(400,270)
\psfig{figure=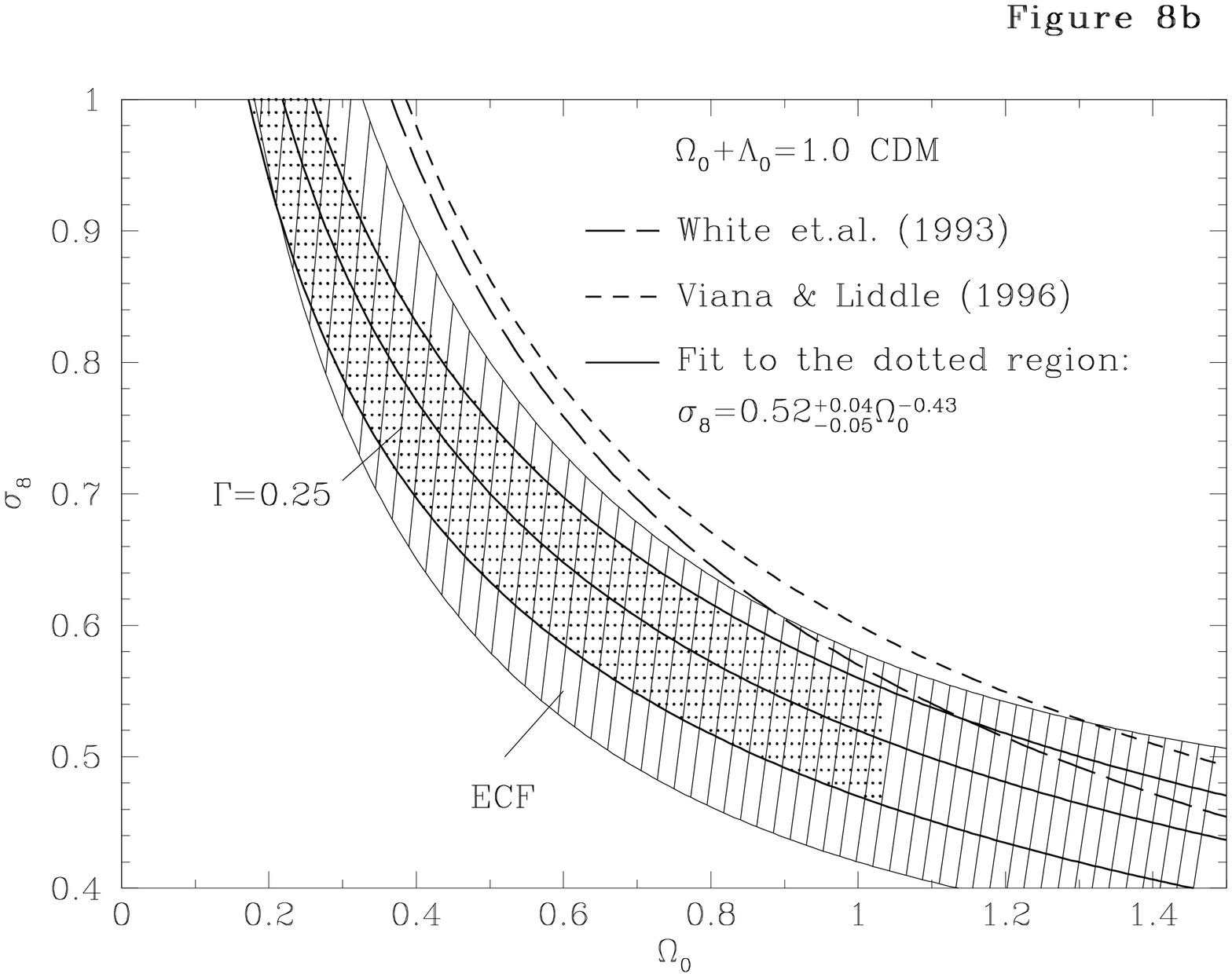,height=10.0cm,width=15.0cm,angle=-90.0}
\end{picture}
\centerline{(8b)}\vspace{0.1in}
\end{figure*}

\begin{figure*}
\centering
\begin{picture}(400,250)
\psfig{figure=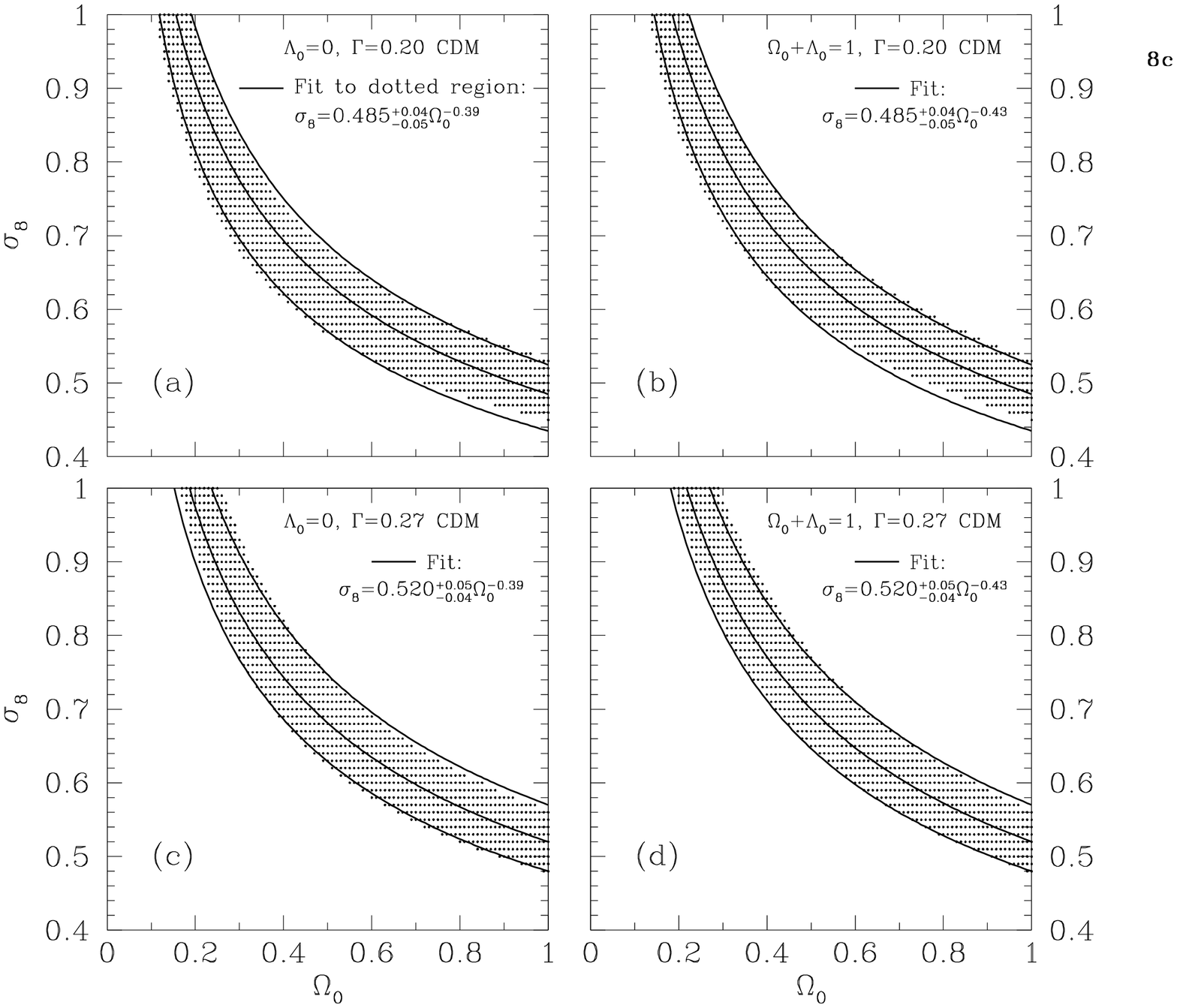,height=10.0cm,width=15.0cm,angle=-90.0}
\end{picture}
\centerline{(8c)}\vspace{0.1in}
\caption{
The constraint on the ($\Omega_0$, $\sigma_8$) plane,
for two cases: $\Gamma=0.25$ and $\Lambda_0=0$ (8a) and
$\Gamma=0.25$ and $\Omega_0+\Lambda_0=1$ (8b).
The dotted region in each figure
represents the permitted regions of parameters
whose models have the cluster mass functions at $z=0$ consistent
with what is observed at $2\sigma$ confidence level.
The hatched region is the fit by Eke \etal (1996)
at $2\sigma$ confidence level
and the short-dashed curve is the fit by Viana \& Liddle (1996).
The long-dashed curve is the fit by White \etal (1993).
The solid curve is the best fit to the dotted region with
the fitting formula being indicated in the figure.
Figure 8c shows the results and fits to them
for $\Gamma=0.20$ and $\Gamma=0.27$.
The best fit of the results, allowing $\Gamma$ to vary within
the indicated range, is presented in equation 36.
}
\end{figure*}

With $\Omega_0$ and $\Lambda_0$ (and the composition of
dark matter and baryonic matter) being fixed, 
the power spectrum transfer function 
of a model becomes completely deterministic.
What is left to be specified is the 
primordial power spectrum index,
$n$ (assuming it is of a power law form).
Figures (7a,b,c,d,e) show 
the constrained two-dimensional parameter space ($\sigma_8,n$)
for the six models listed in Table 2.
The dotted hatched regions are the permitted
space in this two-dimensional parameter plane 
whose cluster mass functions
agree with observations at the $2\sigma$ confidence level.
The solid  hatched regions 
are the constraints provided 
by COBE on large scales,
where the spread in the vertical axis ($\sigma_8$, $\pm 14\%$, $2\sigma$)
is due to the statistical uncertainty of the COBE
measurements (Bunn \& White 1997).
We see that a quite tight constraint
is obtained in the ($n$, $\sigma_8$) plane.
For the four spatially flat models (1,2,5,6) in Table 2,
two cases, with and without
the tensor mode (gravitational wave) contribution
to the CMB fluctuations on COBE scales,
are considered.
We assume the tensor to scalar ratio $T/S=7(1-n)$ 
for $n<1$ (Liddle \& Lyth 1992;
Davis \etal 1992;
Crittenden \etal 1993;
Stewart \& Lyth 1993)
and $T/S=0$ for $n\ge 1$ (Steinhardt 1997, private
communications).
For the remaining two models (open models; models 3,4 in Table 2)
we only consider the case
without the tensor mode contribution;
it turns out to make no difference in the end
whether tensor mode is considered or not,
as the values of
$n$ in the allowed range are greater than unity.
The permitted ranges of $n$, $\sigma_8$ and EP for all the models
are tabulated in Table 3 to clearly show 
their distinctly different ranges for all the models.
However, current observations by COBE on $n$ 
(Gorski \etal 1996;
Bennett \etal 1996;
Hinshaw \etal 1996;
Bunn  \& White 1997)
and by large scale galaxy distribution on EP 
(Wright \etal 1992; or equivalently by $\Gamma$ parameter,
Maddox \etal 1990;
Feldman, Kaiser, \& Peacock 1994;
Peacock, \& Dodd 1994),
as indicated as the last two rows in Table 3,
while significantly constraining models at $1\sigma$ confidence level,
do not yet place $2\sigma$ confidence level
constraint on this set of models.

The values of $n$ and $\sigma_8$ for the central model
in each variant of the CDM model
are shown as solid dots in six panels of Figure 7.
To facilitate further examination of and intercomparison among
the models, the parameters of the six central models
are given in Table 4.
These six models likely bracket all ``viable" CDM models 
of current interest.
This is an attempt to set the context
for future discussions of CDM models that will
share a common standard.

\section{$\sigma_8-\Omega_0$ Relation}

Let us now steer off the main course for a moment 
and examine the $\sigma_8-\Omega_0$ relation
that has been widely utilized (WEF; 
Viana \& Liddle 1996; ECF).
We first note 
that it is very clear from Figure 7 and Table 3
that the shape of the power spectrum plays a significant role.
While we find that
the power spectrum shape dependence of $\sigma_8$
for $\Omega_0=1$ models 
(Models A,B in Table 3) is modest
$\Delta \sigma_8\sim 0.03-0.1$, in agreement with ECF,
the power spectrum shape dependence of $\sigma_8$
for low density models is substantial, being
$\Delta\sigma_8=0.17-0.30$.
Therefore, a more accurate relation of
$\sigma_8-\Omega_0$ cannot be obtained until 
the shape of the power spectrum is more accurately fixed.
The latter is currently unavailable, as indicated by
the last two rows of Table 3.
A tentative solution 
to circumvent this situation
is to choose a shape of the
power spectrum that is deemed to best fit observations of large-scale 
galaxy distribution.
This approach was used by 
WEF, Viana \& Liddle (1996) and ECF,
which will also be adopted here.

Figure 8 shows
the constraint on the ($\Omega_0$, $\sigma_8$) plane,
for two cases:
$\Gamma=0.25$ and $\Lambda_0=0$ (8a) and 
$\Gamma=0.25$ and $\Omega_0+\Lambda_0=1$ (8b).
The dotted region in each figure
represents the permitted regions of parameters 
where models have the cluster mass functions at $z=0$ consistent
with the observations at $2\sigma$ confidence level.
The hatched region is the fit by ECF
at $2\sigma$ confidence level
and the dotted curve is the fit by Viana \& Liddle (1996).
The dashed curve in (8b)
is the fit by WEF.
The solid curve is the best fit to the dotted region.
The general agreements between various studies are very good.
The small differences between the results
of various studies
perhaps reflect the differences in the procedures
of fitting to observational data points,
the small differences in the adopted observations being fit,
and different theoretical approximations used for the models.
Our best fit to the results of GPM to 
the observed mass function of BC, 
for $\Gamma=0.25$, is ($2\sigma$):
$\sigma_8(\Omega_0)=(0.52_{-0.05}^{+0.04})\Omega_0^{-0.39}$
for $\Lambda_0=0$ and
$\sigma_8=(0.52_{-0.05}^{+0.04})\Omega_0^{-0.43}$
for $\Omega_0+\Lambda_0=1$.

Given the sizable allowed range in $\Gamma$ currently constrained
by observation (Peacock \& Dodds 1994),
$\Gamma=0.23_{-0.03}^{+0.04}$ ($2\sigma$),
it is useful to quantify the 
dependence of the above relationships on the shape of the power spectrum.
Figure 8c shows the results and fits to them
for $\Gamma=0.20$ and $\Gamma=0.27$, with the best fits
being ($2\sigma$)
$\sigma_8(\Omega_0)=(0.485_{-0.05}^{+0.04})\Omega_0^{-0.39}$
for $\Lambda_0=0$ and
$\sigma_8=(0.485_{-0.05}^{+0.04})\Omega_0^{-0.43}$
for $\Omega_0+\Lambda_0=1$ for $\Gamma=0.20$,
and 
$\sigma_8(\Omega_0)=(0.52_{-0.04}^{+0.05})\Omega_0^{-0.39}$
for $\Lambda_0=0$ and
$\sigma_8=(0.52_{-0.04}^{+0.05})\Omega_0^{-0.43}$
for $\Omega_0+\Lambda_0=1$ for $\Gamma=0.27$.
The best fit of all the results for CDM-like
models, allowing $\Gamma$ to vary within the indicated range
($0.20-0.27$),
is ($2\sigma$):
\begin{equation}
{\sigma_8(\Omega_0,\Gamma)\hskip -0.1cm =\hskip -0.1cm \cases{&\hskip -0.5cm$[0.50+0.5(\Gamma-0.23)\pm 0.05]\Omega_0^{-0.39}\hskip 0.2cm (\Lambda_0=0$)\cr 
&\hskip -0.52cm$[0.50+0.5(\Gamma-0.23)\pm 0.05]\Omega_0^{-0.43}\hskip 0.2cm (\Omega_0+\Lambda_0=1)$\cr}}.
\end{equation}
\noindent 
This $\sigma_8-\Omega_0$ relation 
best summarizes a constraint on CDM-like models
provided by the observations of zero redshift rich
cluster abundance.

\section{Conclusions}

We have developed and tested a method to compute 
the mass function and correlation function of 
peaks, based on the formalism for Gaussian density field.
We should call this method Gaussian Peak Method (GPM).
The essential new ingredient in this relatively old method 
is a simultaneous
determination of the smoothing window size (to select appropriate peaks)
and the critical peak height of collapse.
A large set of thirty-two N-body simulations are used to test the
accuracy of the method and it is shown that the method
is accurate for all the models tested, which cover 
the parameters space of interest spanned by
$P_k$, $\Omega_0$, $\Lambda_0$ and $\sigma_8$.

The GPM permits economical search of parameter space.
We find that $\sigma_8 - \Omega_0$ relation is somewhat 
dependent upon the shape of the power spectrum.
Normalizing CDM models to the observed local rich cluster
abundance {\it alone},
allowing for the observed uncertainty in the shape parameter
$\Gamma$($\equiv \Omega_0 h$)$=(0.20-0.27)$, gives ($2\sigma$):
\begin{equation}
{\sigma_8(\Omega_0,\Gamma)\hskip -0.1cm =\hskip -0.1cm \cases{&\hskip -0.5cm$[0.50+0.5(\Gamma-0.23)\pm 0.05]\Omega_0^{-0.39}\hskip 0.2cm (\Lambda_0=0$)\cr 
&\hskip -0.52cm$[0.50+0.5(\Gamma-0.23)\pm 0.05]\Omega_0^{-0.43}\hskip 0.2cm (\Omega_0+\Lambda_0=1)$\cr}}.
\end{equation}

Matching both COBE on very large scales
and the abundance of local rich clusters of galaxies 
fixes both the shape ($n$) and amplitude of 
the power spectrum ($\sigma_8$) of any model to about 10\% accuracy.
Consequently, all models become almost completely deterministic.
A set of six CDM models (including cold plus hot dark matter model)
(Table 4), likely bracketing all potentially interesting models,
is advertized together.
This should be viewed as an attempt to set the context
for future discussions on a set of standardized models to 
facilitate comparison of results between workers in the field.

\acknowledgments
The work is supported in part
by grants NAG5-2759 and ASC93-18185.
Discussions with 
Drs. N. Bahcall, J.R. Bond, 
X. Fan, F. Governato, G. Lake, J.P. Ostriker and
D. Scott are gratefully acknowledged.
I thank Dr. F.J. Summers for allowing me to use his P$^3$M
simulation to calibrate the simulation resolution,
and Drs. U. Seljak and M. Zaldarriaga for making
the CMBFAST code available.
Finally, I would like to thank 
George Lake and University of Washington for
the warm hospitality, and financial support
from the NASA HPCC/ESS Program during a visit when
this work was initiated.
The program to compute cluster mass and correlation 
function is available upon request by sending
email to cen@astro.princeton.edu.


\begin{thebibliography}{DUM}
\bibitem[Albrecht \& Steinhardt 1982]{as82} Albrecht, A. \& Steinhardt, P.J. 1982, Phys. Rev. Lett., 48, 1220
\bibitem[Bahcall 1988]{b88} Bahcall, N.A. 1988, ARAA, 26, 631
\bibitem[Bahcall \& Cen 1992]{bc92} Bahcall, N.A., \& Cen, R. 1992, \apj, 398, L81 
\bibitem[Bahcall \& Cen 1993]{bc93} Bahcall, N.A., \& Cen, R. 1993, \apj, 407, L49 (BC) 
\bibitem[Bahcall \& Soneira 1983]{bs83} Bahcall, N.A., \& Soneira, R.M. 1983, \apj, 270, 20 
\bibitem[Bahcall, Fan, \& Cen 1997]{bfc97} Bahcall, N.A., Fan, X., \& Cen, R. 1997, \apj, 485, L53 
\bibitem[Bardeen, Steinhardt, \& Turner 1983]{bst83} Bardeen, J.M., Steinhardt, P.J., \& Turner, M.S. 1983, Phys. Rev. D, 28, 679
\bibitem[Bardeen \etal 1986]{bbks86} Bardeen, J.M., Bond, J.R., Kaiser, N., \& Szalay, A.S. 1986, \apj, 304, 15.
\bibitem[Bardeen, Bond, \& Efstathiou 1987]{bbe87} Bardeen, J.M., Bond, J.R., \& Efstathiou, G. 1987, \apj, 321, 28
\bibitem[Barriola \& Vilenkin 1989]{bv89} Barriola, M. \& Vilenkin, A. 1989, Phys. Rev. Lett., 63, 341
\bibitem[Bartlett \& Silk 1993]{bs93} Bartlett, J.G., \& Silk, J. 1993, \apj, 407, L45
\bibitem[Baugh, Gaztanaga, \& Efstathiou 1995]{bge95} Baugh, C.M., Gaztanaga, E., \& Efstathiou, G. 1995, \mnras, 274, 1049
\bibitem[Bennett \etal 1996]{be96} Bennett, C.L., \etal 1996, ApJ, 464, L1
\bibitem[Bennett \& Rhie 1990]{br90} Bennett, D.P., \& Rhie, S.H. 1990, Phys. Rev. Lett., 65, 1709
\bibitem[Bond \& Couchman 1988]{bc88} Bond, J.R., \& Couchman, H.M.P. 1988,
in Proc. Second Canadian Conference on General Relativity \& Relativistic
Astrophysics, eds. A. Coley, C.C. Dyer, \& B.O.J. Tupper (Singapore: World Scientific), 385
\bibitem[Borgani \etal 1995]{betal95} Borgani, S., Plionis, M., Coles, P., \& Moscardini, L. 1995, \mnras, 277, 1191
\bibitem[Brieu, Summers, \& Ostriker 1995]{bso95} Brieu, P., Summers, F.J., \& Ostriker, J.P. 1995, ApJ, 453, 566
\bibitem[Bunn \& White 1997]{bw97} Bunn, E.F., \& White, M. 1997, \apj, 480, 6
\bibitem[Bucher, Goldhaber, \& Turok 1995]{bgt95} Bucher, Goldhaber, \& Turok, N. 1995, preprint
\bibitem[Burles \& Tytler 1997]{bt97} Burles, S., \& WTytler, D. 1997, preprint,astro-ph/9712108
\bibitem[Carlberg \etal 1996]{cyeag96}Carlberg., R.G., Yee, H.K.C., Ellingson, E., Abraham, R., Gravel, P., Morris, S.M., \& Pritchet, C.J., 1996, ApJ, 462, 32
\bibitem[Postman \etal 1996]{plgosc96}Postman, M., Lubin, L., Gunn, J.E., Oke, J.G., Schneider, D.P., \& Christensen, J.A., 1996, ApJ, 111, 615
\bibitem[Cen 1997]{c97} Cen, R. 1997b, \apj, 485, 39
\bibitem[Cen 1998]{c98} Cen, R. 1998, \apj, in press
\bibitem[Cen, Bahcall, \& Gramann 1994]{cbg94} Cen, R., Bahcall, N.A., \& Gramann, M. 1994, \apj, 437, L51
\bibitem[Colless 1998]{col98} Colless, M. 1998, preprint, astro-ph/9804079
\bibitem[Colley 1997]{colley97} Colley, W.N. 1997, \apj, 489, 471
\bibitem[Colley, Gott, \& Park 1996]{cgp96} Colley, W.N., Gott, J.R., III, \& Park, C. 1996, \mnras, 281, L82
\bibitem[Crittenden \etal 1993]{cetal93} Crittenden, R., Bond, J.R., Davis, R.L., Efstathiou, G., \& Steinhardt, P.J. 1993, Phys. Rev. Lett., 71, 324
\bibitem[Croft \& Efstathiou 1994]{ce94} Croft, R.A.C., \& Efstathiou, G.  1994, \mnras, 267, 390
\bibitem[Croft \etal 1997]{cdesm97} Croft, R.A.C., Dalton, G.B., Efstathiou, G., Sutherland, W.J., \& Maddox, S.J. 1997, \mnras, 291, 305
\bibitem[Dalton \etal 1992]{dems94} Dalton, G.B., Efstathiou, G., Maddox, S.J., \& Sutherland, W.J. 1992, \apj, 390, L1
\bibitem[Dalton \etal 1994]{dcesmd94} Dalton, G.B., Croft, R.A.C., Efstathiou, G., Sutherland, W.J., Maddox, S.J., \& Davis, M. 1994, MNRAS, 271, L47
\bibitem[Danos \& Pen 1998]{dp98} Danos, R., \& Pen, U. 1998, preprint, astro-ph/9803058
\bibitem[Davis \& Peebles 1983]{dp83} Davis, M., \& Peebles, P.J.E. 1983, ApJ, 267, 465
\bibitem[Davis \etal 1992]{d92} Davis, R.L., Hodges, H.M., Smoot, G.F., Steinhardt, P.J., Turner, M.S. 1992, Phys. Rev. Lett., 69, 1856; erratum, 70, 1733
\bibitem[Dressler 1980]{d80} Dressler, A. 1980, \apjs, 42, 565
\bibitem[Dressler \& Shectman 1988]{ds88} Dressler, A., \& Shectman, S.A. 1988, AJ, 95, 985 
\bibitem[Efstathiou \& Rees 1988]{er88} Efstathiou, G., \& Rees, M. 1988, \mnras, 230, 5
\bibitem[Eke, Cole, \& Frenk 1996]{ecf96} Eke, V.R., Cole, S., \& Frenk, C.S. 1996, \mnras, 282, 263 (ECF)
\bibitem[Einasto \etal 1997a]{e97a} Einasto, J., Einasto, M., Gottlober, S., Muller, V., Saar, V., Starobinsky, A.A., Tago, E., Tucker, D., Andernach, H., \& Frisch, P. 1997, Nature, 385, 139
\bibitem[Einasto \etal 1997b]{e97b} Einasto, J., Einasto, M., Frisch, P., Gottlober, S., Muller, V., Saar, V., Starobinsky, A.A., Tago, E., Tucker, D., \& Andernach, H. 1997, \mnras, 289, 801
\bibitem[Einasto \etal 1997c]{e97c} Einasto, J., Einasto, M., Frisch, P., Gottlober, S., Muller, V., Saar, V., Starobinsky, \& E., Tucker, D. 1997, \mnras, 289, 813
\bibitem[Feldman, Kaiser, \& Peacock 1994]{fnp94} Feldman, H.A., Kaiser, N., \& Peacock, J.A. 1994, \apj, 437, 56
\bibitem[Fischer \etal 1997]{fgrt97} Fischer, P., Bernstein, G., Rhee, G., \& Tyson, J.A. 1997, AJ, 113, 521 
\bibitem[Geller \& Beers 1982]{gb82} Geller, M.J, \& Beers, T.C. 1982, Publ. Ast
ron. Soc. Pac., 94, 421
\bibitem[Goldberg \& Strauss 1998]{gs98} Goldberg, D.M., \& Strauss, M.A. 1998, \apj, 495, 29
\bibitem[Gorski \etal 1996]{g96} Gorski, K.M., Banday, A.J., Bennett, C.L., Hinshaw, G., Kogut, A., Smoot, G.F., \& Wright, E.L. 1996, ApJ, 464, L11 
\bibitem[Gott 1982]{g82} Gott, J.R., III 1982, \nat, 295, 304
\bibitem[Gunn \& Gott 1972]{gg72} Gunn, J.E., \&  Gott, J.R  1972, ApJ, 176, 1.
\bibitem[Guth \& Pi 1982]{gp82} Guth, A.H., \& Pi, S.-Y. 1982, Phys. Rev. Lett. 49, 1110
\bibitem[Henry \& Arnaud 1991]{ha91} Henry, J.P., \& Arnaud, K.A. 1991, \apj, 372, 410
\bibitem[Hinshaw \etal 1996]{h96} Hinshaw, G., \etal 1996, ApJ, 464, L17
\bibitem[Holtzman \& Primack 1993]{hp93} Holtzman, J.A., \& Primack, J.R. 1993, \apj, 405, 428
\bibitem[Kaiser 1984]{k84} Kaiser, N. 1984, ? 
\bibitem[Kaiser \& Davis 1985]{kd85} Kaiser, N., \& Davis, M. 1985, \apj, 297, 365
\bibitem[Klypin \& Kopylov 1983]{kk83} Klypin, A.A., \& Kopylov, A.I. 1983, Soviet Astr. Letters, 9, 41
\bibitem[Klypin \etal 1995]{ketal95} Klypin, A., Borgani, S., Holtzman, J., \& Primack, J. 1995, \apj, 444, 1
\bibitem[Kogut \etal 1996]{k96} Kogut, A., Banday, A.J., Bennett, C.L., Gorski, K.M., Hinshaw, G., \& Smoot, G.F. 1996, ApJ, 464, L29
\bibitem[Lacey \& Cole 1993]{lc93} Lacey, C., \& Cole, S. 1993, MNRAS, 262, 627
\bibitem[Liddle \& Lyth 1992]{ll92} Liddle, A.R., \&  Lyth, D.H. 1992, Phys. Lett., B 291, 391
\bibitem[Linde 1982]{l82} Linde, A. 1982, Phys. Lett, 108B, 389
\bibitem[Lowenthal \etal 1997]{l97} Lowenthal, J.D., Koo, D.C., Guzman, R., Gallego, J., Phillips, A.C., Faber, S.M., Vogt, N.P., Illingworth, G.D., \& Gronwall, C. 1997, \apj, 481, 673
\bibitem[Lu \etal 1997]{lsb97} Lu, L., Sargent, W.L.W., \& Barlow, T.A. 1997, \apj, 484, 131
\bibitem[Lubin \etal 1996]{lcbo96} Lubin, L.M., Cen, R., Bahcall, N.A., \& Ostriker, J.P. 1996, \apj, 460, 10
\bibitem[Luppino \& Gioia 1995]{lg95} Luppino, G.A., \& Gioia, I.M. 1995, \apj, 445, 77 (LG95)
\bibitem[Maddox \etal 1990]{mesl90} Maddox, S.J., Efstathiou, G., Sutherland, W.J., \& Loveday, J. 1990, \mnras, 242, 43p
\bibitem[Mann, Heavens, \& Peacock 1993]{mhp93} Mann, R.G., Heaven, A.F., \& Peacock, J.A. 1993, \mnras, 263, 798
\bibitem[More, Heavens, \& Peacock 1986]{mhp86} More, J.G., Heavens, A.F., \& Peacock, J.A. 1986, \mnras, 220, 189
\bibitem[Navarro, Frenk, \& White 1996]{nfw96} Navarro, J.F., Frenk, C.S., \& White, S.D.M. 1996, ApJ, 462, 563
\bibitem[Nichol \etal 1992]{ncgl92} Nichol, R.C., Collins, C.A., Guzzo, L., \& Lumsden, S.L. 1992, MNRAS, 255, 21p
\bibitem[Oukbir \& Blanchard 1992]{ob92} Oukbir, J., \& Blanchard, A. 1992, A\& A, 262, L21
\bibitem[Peacock, \& Dodd 1994]{pd94} Peacock, J.A., \& Dodds, S.J. 1994, \mnras, 267, 1020
\bibitem[Peacock \& Heavens 1985]{ph85} Peacock, J.A., \& Heavens, A.F. 1985, MNRAS, 217, 805
\bibitem[Peebles 1980]{p80} Peebles, P.J.E. 1980, The Large-Scale Structure of the Universe (Princeton: Princeton University Press)
\bibitem[Peebles, Daly, \& Juszkiewicz 1989]{pdj89} Peebles, P.J.E., Daly, R.A., \& Juszkiewicz, R. 1989, \apj, 347, 563
\bibitem[Pen 1996]{p96} Pen, U.-L. 1996, preprint, astro-ph/9610147
\bibitem[Postman, Huchra, \& Geller 1992]{phg92} Postman, M., Huchra, J.P., \&  Geller, M.J. 1992, ApJ, 384, 404
\bibitem[Press \& Schechter 1974]{ps74} Press, W.H., \& Schechter, P.L. 1974, \apj, 215, 703
\bibitem[Protogeros \& Weinberg 1997]{pzw97} Protogeros, Z.A.M., \& Weinberg, D.H. 1997, \apj, 489, 457
\bibitem[Richstone, Leob, \& Turner 1992]{rlt92} Richstone, D., Loeb, A., \& Turner, E.L. 1992, \apj, 393, 477
\bibitem[Romer \etal 1994]{rcbcemv94} Romer, A.K., Collins, C.A., Bohringer, H.,Cruddace, R.G., Ebeling, H., MacGillivray, H.T., \& Voges, W. 1994, Nature, 372, 75
\bibitem[Salaris, Degl'Innocenti, \& Weiss 1997]{sdw97} Salaris, M., Degl'Innocenti, S., \& Weiss, A. 1997, \apj, 479, 665
\bibitem[Smoot \etal 1992]{s92} Smoot, G.F. \etal 1992, \apj, 396, L1.
\bibitem[Stewart \& Lyth 1993]{sl93} Stewart, E., \& Lyth, D.H. 1993, Phys., Lett. B, 302, 171
\bibitem[Strauss \& Willick 1995]{sw95} Strauss, M.A., \& Willick, J.A. 1995, Phys. Repor, 261, 271
\bibitem[Strauss \etal 1995]{s95} Strauss, M.A., Cen, R., Ostriker, J.P., Lauer, T.R., \& Postman, M. 1995, \apj, 444, 507
\bibitem[Suto, Cen, \& Ostriker 1992]{sco97b} Suto, Y., Cen, R., \& Ostriker, J.P. 1992, 395, 1
\bibitem[Trimble 1997]{t97} Trimble, V. 1997, ``The Extragalactic Distance Scale" ed.  M. Livio, M. Donahue \& N. Panagia, p407
\bibitem[Turok 1989]{t89} Turok, N. 1989, Phys. Rev. Lett., 63, 2625
\bibitem[Viana \& Liddle 1996]{vl96} Viana, P.T.P, \& Liddle, A.R. 1996, \mnras, 281, 323
\bibitem[Vilenkin 1981]{v81} Vilenkin, A. 1981, Phys. Rev. Lett., 46, 1169
\bibitem[Vilenkin 1985]{v85} Vilenkin, A. 1985, Phys. Rep., 121, 263
\bibitem[Vogeley \etal 1994]{v94} Vogeley, M.S., Park, C., Geller, M.J., Huchra, J.P., \& Gott, J.R., III 1994, \apj, 420, 525
\bibitem[Walker \etal 1991]{w91} Walker, T.P. et al. 1991, \apj, 376, 51
\bibitem[West, Oemler, \& Dekel 1988]{wod88} West, M.J., Oemler, A., Jr., \& Dekel, A. 1988, \apj, 327, 1
\bibitem[White \etal 1993a]{wef93} White, S.D.M., Efstathiou, G., \& Frenk, C.S. 1993a, \mnras, 262, 1023 (WEF)
\bibitem[White \etal 1993]{wnef93} White, S.D.M., Navarro, J.F., Evrard, A.E., \& Frenk, C.S. 1993, \nat, 366, 6454
\bibitem[Wright \etal 1992]{w92} Wright, E.L., \etal 1992, ApJ, 396, L13
\bibitem[Xu 1995]{x95} Xu, G. 1995, \apjs, 98, 355
\bibitem[Zel'dovich 1980]{z80} Zel'dovich, Ya. B. 1980, MNRAS, 192, 663
\end{thebibliography}
\end{document}